\documentclass[preprint]{elsarticle}
\usepackage{geometry}
\usepackage{array}
\usepackage{amsmath}
\usepackage{makecell}
\usepackage{cleveref}
\usepackage{graphicx}
\usepackage{subcaption}
\usepackage{mathrsfs}
\usepackage{xcolor}
\usepackage{ulem}

\biboptions{sort&compress}
\Crefname{figure}{Fig.}{Figs.}
\Crefname{equation}{Eq.}{Eqs.}

\newcommand{\V}{_{\text{V}_{\text{O}}^{..}}}
\newcommand{\Vb}{_{\text{V}_{\text{O}}^{..},\text{b}}}
\newcommand{\Vc}{_{\text{V}_{\text{O}}^{..},\text{c}}}
\newcommand{\dop}{_{\text{dop}}}
\newcommand{\dopb}{_{\text{dop},\text{b}}}
\newcommand{\dopc}{_{\text{dop},\text{c}}}

\begin{document}
\begin{frontmatter}

\title{A defect-chemistry-informed phase-field model of grain growth in oxide ceramics: application to Fe-doped SrTiO$_3$}


\author[a]{Kai Wang}
\author[b]{Roger A. De Souza}
\author[a]{Xiang-Long Peng}
\author[c]{Rotraut Merkle}
\author[d]{Wolfgang Rheinheimer}
\author[e]{Karsten Albe}
\author[a]{Bai-Xiang Xu\corref{cor1}}

\cortext[cor1]{Corresponding author}

\address[a]{Mechanics of Functional Materials Division, Institute of Materials Science, Technische Universit\"at Darmstadt, Darmstadt, 64287, Germany}
\address[b]{Institute of Physical Chemistry, RWTH Aachen University, Aachen, 52056, Germany}
\address[c]{Max Planck Institute for Solid State Research, Heisenbergstra\ss e 1, Stuttgart, 70569, Germany}
\address[d]{University of Stuttgart, Institute for Manufacturing Technologies of Ceramic Components and Composites (IFKB), Stuttgart, 70569, Germany}
\address[e]{Materials Modelling Division, Institute of Materials Science, Technische Universit\"at Darmstadt, Darmstadt, 64287, Germany}


\begin{abstract}
Dopants can significantly affect the properties of oxide ceramics through their impact on the property-determined microstructure characteristics such as grain boundary (GB) segregation, space charge layer formation in the GB vicinity, and the grain growth deviating from normal patterns. To support the rational design of oxide ceramics, we propose a defect-chemistry-informed phase-field grain growth model to simulate the microstructure evolution of oxide ceramics. 
{It fully respects the defect-chemistry theory by accounting for the distinct segregation energies and available site densities of charged point defects (oxygen vacancies and acceptor dopants) in both the grain interior and boundaries,} and it considers the competing kinetics of defect diffusion and GB movement.
{The proposed phase-field model is benchmarked against well-known  bicrystal models, including the Mott-Schottky and Gouy-Chapman models. Various simulation results are presented to reveal the effect of different defect-chemistry parameters on the space charge layer formation and key microstructural aspects.} In particular, simulation results confirm that the solute drag effect alone can lead to {skewed grain size distribution that do not follow the log-normal distribution}, without any contribution from grain misorientation and other anisotropy. Interestingly,  simulations also demonstrate that grain boundary potentials can vary substantially: grain boundaries of larger grains tend to have lower potentials than those of smaller grains.
Such heterogeneous grain boundary potential distribution may inspire a new material optimization strategy through microstructure design. This study provides a comprehensive framework for defect-chemistry-consistent investigations of microstructure evolution in polycrystalline oxide ceramics, offering fundamental insights into microscopic processes during critical manufacturing stages.
\end{abstract}

\begin{graphicalabstract}
\includegraphics{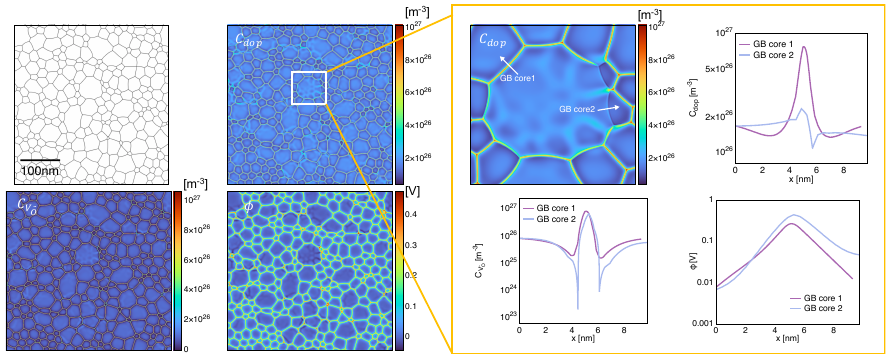}
\end{graphicalabstract}



\end{frontmatter}

\section{Introduction} \label{introduction}

In recent decades, oxide ceramics, such as SrTiO$_3$ (STO) and BaTiO$_3$ (BTO), have undergone tremendous development and are increasingly crucial in applications, such as capacitors 
 \cite{lee2011atomic,hou2017ultrahigh}, actuators \cite{kursumovic2013new,baek2013epitaxial}, sensors  \cite{wang2022stretchable,chan2014highly},  memristors  \cite{muenstermann2010coexistence, hu2013ferroelectric} and electrolytes in fuel cells 
 \cite{ishihara1994doped, kreuer2003proton, li2014family},   
 {Oxide ceramics can be fabricated via sintering or deposition \cite{mishra2022ultra,reavley2022ultrafast,sayer1990ceramic,li2003two}.}
 When manufactured through sintering, oxide ceramics exhibit a polycrystalline microstructure, with their macroscopic properties heavily influenced by the synthesis conditions \cite{chaim2008sintering}. 
To tailor {the physical and chemical} properties of oxide ceramics, heterovalent substitution (doping) is an effective method for introducing additional charged defects through ionic or electronic compensation during synthesis. The state-of-the-art designs are mostly empirical, since knowledge of the relationships between doping strategies, processing and the resultant microstructure is limited.
The influential mechanisms of doping on microstructure in oxide ceramics are manifold. First of all, the addition of doping elements in the parent material systems triggers defect chemistry reactions and results in multiple types of charged point defects or defect complexes, which interact with the grain boundary (GB) chemically, electrically and elastically and form space charge layers (SCLs) around GB. 
More specifically, the equilibrium concentrations of the point defects within the bulk are thermodynamically determined through defect chemistry \cite{maier2023physical, smyth2000defect}. The scenario near inside the GB core and in the GB vicinity are different.
{In undoped and acceptor doped perovskites such as Fe-doped STO}, the GB core is positively charged with excess oxygen vacancies,  $\text{V}_{\text{O}}^{\text{..}}$ in Kr\"oger-Vink notation \cite{kroger1956relations}, stemming from the negative formation energy difference of oxygen vacancy between GB core and bulk, i.e. $\Delta g\V = \mu^0\Vc - \mu^0\Vb < 0$.
To compensate the positively charged GB core, negatively charged regions form adjacent to it, called the space charge layers (SCLs). 

\begin{figure}
    \centering
\includegraphics[width=0.8\linewidth]{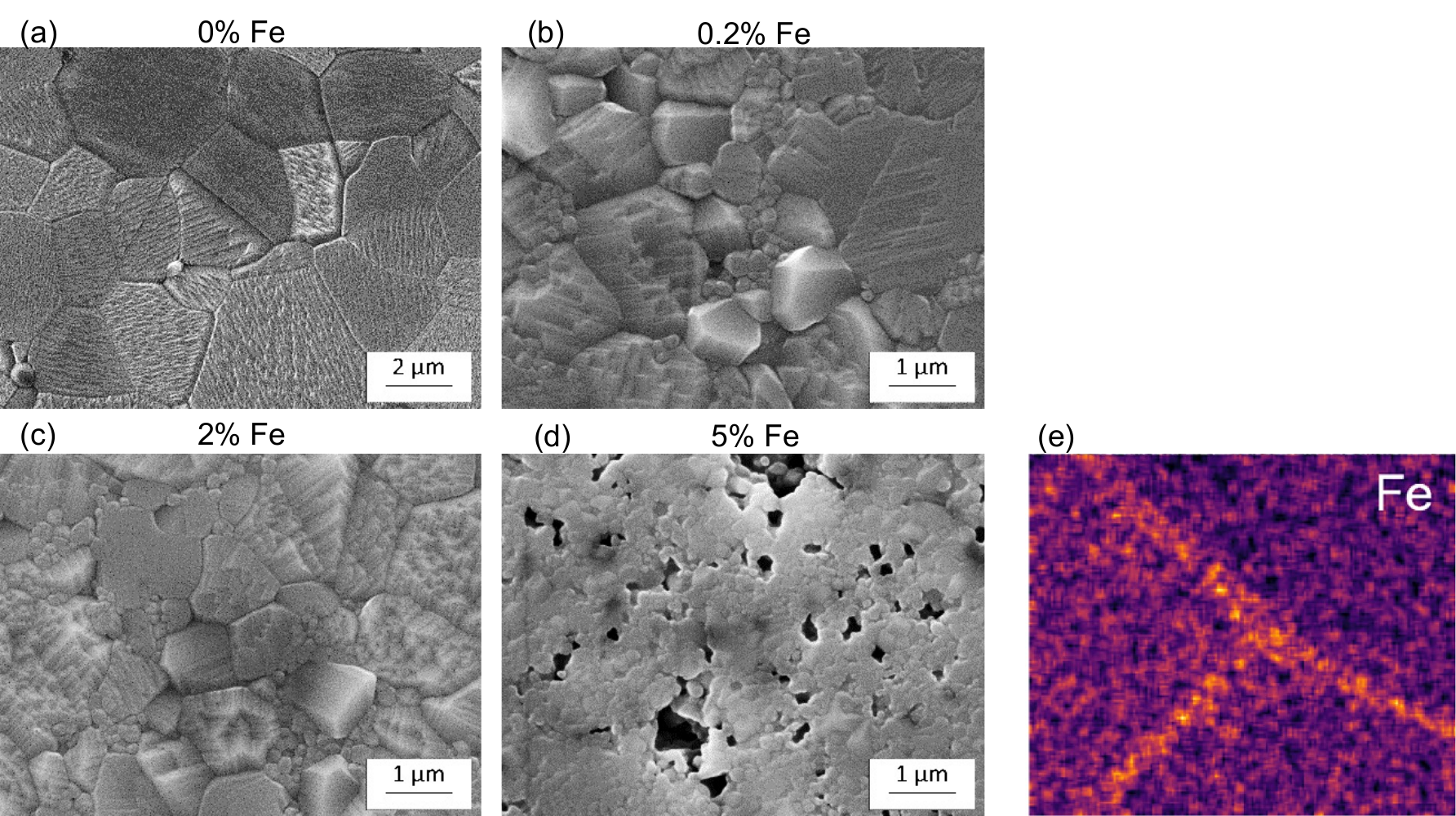}
    \caption{Microstructures in undoped (a) and Fe-doped STO (b-d) after heating to 1300$^\circ \text{C}$ for 10h. With increasing Fe concentration, less grain growth occurs. A fraction of very small grains appears in the microstructures for 0.2\% Fe. For 2\% Fe, (c), more such small grains are evident, until for 5\% Fe, all grains remain small. Note the different scale in (a) \cite{zahler2023grain}. (e) Fe segregation to the grain boundaries as evident by STEM-EDS \cite{jennings2024grain}. Sources:  10.1016/j.jeurceramsoc.2022.11.074, 10.1016/j.actamat.2024.119941}
    \label{fig_exp}
\end{figure}

Sharp interface models have been used to {describe} the SCL and the related grain boundary potential. Among them are the Mott-Schottky (MS) model \cite{mott1939theory, mott1938note, schottky1939halbleitertheorie} and the Gouy-Chapman (GC) model \cite{guoy1910constitution, chapman1913li}.
The MS model assumes {that the dopant concentration is constant}, and only the oxygen vacancies {are able to redistribute, in order to form the SCLs}.
In contrast, the GC model {has both defects as mobile and thus allows both to contribute to charge compensation}. While positively charged oxygen vacancies are depleted in the case of a positive core, acceptor dopants are depleted in the case of a positive core within the SCL and in the GB core {due to their negative charge [\Cref{fig_exp}(e)]}. 
The spatial inhomogeneity introduced by SCL formation exerts a profound impact on the transport dynamics of charge species. Evidence shows that cation diffusion parallel to the GB is expedited \cite{parras2020grain, savzinas2017134ba}, whereas charge transport across the GB is significantly hindered \cite{waser2000grain}. 

SCLs determine not only the GB local properties of the synthesized material, but also impact the grain growth dynamics already during the synthesis through the well-known solute drag effect \cite{cahn1962impurity}. Grain growth in oxide ceramics, driven by {a reduction in surface area} and controlled by the diffusion of charged species, plays a crucial role in determining the final grain size distribution and the concentration distributions of various point defects.
Segregation in a moving GB core necessitates additional free energy dissipation and results in the solute drag effect \cite{cahn1962impurity, hillert1976treatment}. This effect strongly affects the kinetics of grain growth and can independently trigger abnormal grain growth even in the absence of any texture or pinning sites \cite{kim2008grain}. Dopant segregation to GB core has been experimentally evidenced, especially at high temperatures \cite{zahler2023grain, jennings2024grain}. Experiments also show that even 0.2\% amount of Fe in STO can pin a certain portion of GBs [\Cref{fig_exp}(b)]. This trend increases at 2\% Fe [\Cref{fig_exp}(c)]. When increasing the dopant concentration to 5\% [\Cref{fig_exp}(d)], the grain growth is completely suppressed \cite{zahler2023grain, rheinheimer2016grain}. In addition, the grain growth coefficient is found to follow a non-Arrhenius behavior between $1350^\circ$C and $1425^\circ$C. Two types of GB core with different mobilities have been proposed to explain the non-Arrhenius behavior \cite{rheinheimer2015non, zahler2023grain}. 
{Despite various pieces of evidence supporting the impact of SCLs on GB properties and grain growth dynamics, experimental characterization of the SCL remains a challenge \cite{fleig2002grain}. }
The complexity arises from the inherent structural and compositional heterogeneities present in polycrystalline oxide ceramics. One of the primary obstacles in this characterization is the limitation of in-situ transmission electron microscopy (TEM) investigations. Such studies are typically unfeasible due to the high sintering temperatures required for these ceramics, which exceed the operational limits of most TEM setups.

It is hence desirable to employ theoretical models and numerical simulations to evaluate SCLs and abnormal grain growth. 
{Kliewer and Koehler derived the local density of charge carriers in the presence of electrostatic potentials from a global approach \cite{kliewer1965space}. Through minimizing the free energy of the entire crystal, the equations for the densities of various defects are derived in the type of Boltzmann distribution. Blakely and Macdonald extended the global thermodynamic approach with the consideration of an exhaustible number of defect sites and derived Fermi-Dirac-like distribution functions of defects \cite{blakely1973space, mukhopadhyay1991ionic, macdonald1980interfacial}. Jamnik employed a local thermodynamic formalism, in which the interface region was divided into a core and a space charge region with an assumption that the material constants, i.e. the standard chemical potentials and mobilities of defects, behave as step functions \cite{jamnik1995interfaces}. 
De Souza extended the local thermodynamic approach to abrupt GB$|$SCL model  based on defect chemistry in acceptor-doped {oxide systems} at equilibrium state \cite{de2009formation}}. The impacts of different charge species, the formation energy difference of oxygen vacancy and dopant between GB core and bulk phase, {the number of available sites of oxygen vacancy and dopants}, temperature and partial pressure of oxygen on grain boundary potential are comprehensively studied in a STO bicrystal with planar GB core \cite{de2009formation}. {Then, the abrupt GB$|$SCL model was implemented to reproduce the SCL in STO to investigate the impact of negative electrode on the defects distribution, GB potential and grain growth kinetics \cite{rheinheimer2019grain}. In addition, this model was also employed to investigate the SCL formation in fluorite-type oxide \cite{parras2020grainJACS, tong2020analyzing, parras2020grain}.}

{However, applying the abrupt GB$|$SCL model  to polycrystalline sintering scenarios is challenging. The complex microstructure, intricate GB distribution and GB kinetics affected by solute drag effects hinder the practical use of the sharp interface methods.
Under this circumstance, the phase-field method with a diffuse interface emerges as a powerful tool to solve different moving boundary problems \cite{wang2020modeling,wang2021quantitative, hu2024data}. Kobayashi et al. developed the Kobayashi-Warren-Carter (KWC) phase-field model to consider crystallographic orientation during grain growth \cite{kobayashi2000continuum}. Moelans et al. proposed a homogeneous multi-well potential in the free energy density functional to obtain the quantitative relations between phase-field model parameters and GB properties \cite{moelans2008quantitative}.  Additionally, there are also a series of phase-field sintering models for  non-isothermal sintering \cite{yang20193d,yang2020investigation,yang2022diffuse,oyedeji2023variational}.}
{Several researchers have developed phase-field models to capture solute drag effects during grain growth processes. The first phase-field simulation was performed by Fan et al. \cite{fan1999computer}. Their model revealed that solute drag introduces a nonlinear relationship between GB velocity and driving force and influences the growth exponent in polycrystalline systems.  Cha et al. proposed a phase-field model that treats the GB as a distinct phase and studied the impact of solute drag on migrating grain boundaries in a binary alloy system \cite{cha2002phase}. The effect of solute drag is automatically incorporated into the model and it is capable of reproducing the equilibrium solute segregation and the free energy dissipation. Gr\"onhagen et al. introduced a concentration dependency in the height of the double-well potential in the Gibbs-energy expression and verified the possibility of modeling solute drag on a moving boundary \cite{gronhagen2007grain}.
This model was then adopted by Kim \cite{kim2008grain} and Li \cite{li2009phase}. Kim et. al investigated the grain growth in association with GB segregation in two-dimensional polycrystalline systems \cite{kim2008grain}. They concluded abnormal grain growth can be triggered by the solute drag effect.  
Li et al. observed that the solute concentration at the moving GB may increase with increasing velocity and becomes larger than the equilibrium value, which is unexpectedly predicted by solute drag theory \cite{li2009phase}. They also found non-linear relationship between the curvature driving force and the GB velocity can exist not only in the transition regime of the velocity, but also in the low velocity regime.
Greenwood et al. studied the atomic scale interaction of solute with a migrating GB by using a phase-field crystal model \cite{greenwood2012phase}. They developed a new formalism to allow for the application of an external driving pressure for the growth of one grain. Extending from grain growth studies,  phase-field models for solid-state sintering in the initial and final stage considering various diffusion paths are proposed by Wang et al. \cite{wang2006computer}, Kumar
et al. \cite{kumar2010phase}, H\"otzer et al. \cite{hotzer2019phase} and  Rehn et al.  \cite{rehn2019phase}. Rheinheimer et. al employed a multi-phase-field model to simulate the abnormal grain growth in STO ceramics by introducing two types of GB mobilities \cite{rheinheimer2019non}. However, the electrostatic effects related SCL is not considered. }

{In order to consider the electrostatic contribution in the phase-field model, Guyer et al. developed a phase-field model to capture the charge distribution in an equilibrium double layer at the electrochemical interface with mass and volume constraints, Poisson’s equation, ideal solution thermodynamics, and a simple description of the competing energies in the interface \cite{guyer2004equilibrium}. By using the same model, Guyer et al. also explored its kinetic behavior for electrodeposition and electrodissolution conditions \cite{guyer2004kinetic}. They demonstrated the ohmic conduction in the electrode and ionic conduction in the electrolyte and found the nonlinear relationship between current and overpotential as predicted by Butler-Volmer equation. Bishop et al. formulated  a Helmholtz free energy functional with the electrostatic contribution and studied the phase separation of charged species by spinodal decomposition \cite{bishop2003effect}. Garc\'ia et al.  incorporated the laws of thermodynamics and Maxwell's equations and proposed a theoretical framework including thermodynamically consistent equilibrium equations and kinetic driving forces to describe the time evolution for electrically and magnetically active materials \cite{garcia2004thermodynamically}. Lund et al. proposed a generalized framework that integrates the free energy contributions of thermochemical, structural, mechanical, and electrical fields \cite{lund2021thermodynamically}. They calculated the SCL in gadolinium-doped cerium oxide with high dopant concentration and predicted the macroscopic ionic conductivity with different segregation energy of oxygen vacancy.
}

{Aiming to investigate the sintering process with electrostatics, multi-physics phase-field model needs to be developed. Vikrant et al. developed a phase-field model including the electro-chemo-mechanical effects to physically describe the equilibrium and transport properties of charged interfaces in ion-conducting solids \cite{vikrant2018charged}. They simulated gadolinium-doped cerium oxide and investigated the influence of gadolinia substitution concentration on the SCL. Then, Vikrant et al. extended this model to understand the effects of the intrinsic and extrinsic ionic species and point defects on the structural and electrochemical stability of grain boundaries in polycrystalline ceramics \cite{vikrant2019charged}. They studied the impacts of different crystallographic misorientations on SCL in the cubic Yttria Stabilized Zirconia.
{Recently, Vikrant et al. developed a phase-field model based on KWC grain growth model and investigate the SCL formation at equilibrium and quasi-equilibrium state with the consideration of the GB misorientation and solute drag effect \cite{vikrant2020electrochemical}.} They considered the negative segregation formation energy of oxygen vacancies for SCL formation (and unfortunately treated oxygen vacancies and cation dopants as sharing the same crystallographic site). Then, this model was applied to explain the abnormal grain growth with a bimodal pattern as observed in experiments \cite{vikrant2020electrochemicallyActa}. In addition to free energy based phase-field model, Aagesen et al. developed a grand potential-based phase-field model and included includes the effects of charged vacancies and the associated interactions between internal and applied electric fields \cite{aagesen2024electrochemical}.  {In oxide ceramics such as Fe-doped STO, Fe atoms substitute for  Ti$^{4+}$ ions on the B-site of the perovskite sublattice, while charge compensation typically occurs via the formation of oxygen vacanies on the anion site. As a result, in the bulk phase, the number of available sites for an acceptor dopant corresponds to the number density of B-sites, which is $1/a^3$, where $a$ is the lattice parameter. The number of available sites indicates the maximum number of sites occupied by defects. In contrast, the number of available sites for an oxygen vacancy is three times higher, i.e. $3/a^3$.  Even more important, B-site dopant and oxygen vacancy belong to different sub-lattices and thus have to be accounted for separately, in order to correctly represent the respective configurational entropy \cite{maier2023physical}.  In addition, the local structure within the GB core differs from that of the bulk. These structural differences can significantly alter the number of available sites for both dopants and oxygen vacancy. Therefore, it is necessary to treat the number of available sites for oxygen vacancies and dopants in the GB core as distinct from those in the bulk and these quantities play a critical role in determining the SCL. Variations in the number of available sites for oxygen vacancy and dopant give rise to diverse SCL formation behaviors \cite{de2009formation}.  Although Vikrant and Aagesen performed landmark studies on the space charge layer formation and grain microstructure evolution during the sintering process, the accurate coupling of these defect chemistry parameters into phase-field simulations was not addressed.}} 


{Therefore, defect-chemistry consistent SCLs are characterized by a number of inequivalent defects, each with its unique segregation energy and number of available sites. Some GB cores exhibit not only a negative segregation formation energy for oxygen vacancies ($\Delta_\text{seg} g\V<0$), but also a combined negative segregation formation energy for oxygen vacancies and acceptor cations ($\Delta_\text{seg} g\V<0$ and $\Delta_\text{seg} g\dop<0$) \cite{usler2024point}. Meanwhile, different defects in the bulk and GB core require different numbers of available sites \cite{usler2024point, usler2024space, de2019effect}. Moreover,  the dominant effect leading to abnormal grain growth during the sintering process in oxide electroceramics remains debatable. The combined effects of GB misorientation and solute drag have been explored in Fe-doped STO \cite{vikrant2020electrochemical, vikrant2020electrochemicallyActa}. Nevertheless, an investigation of GB energy anisotropy by the GB grooving technique indicates that the reduction of GB energy due to anisotropy can only lead to moderate (about 15\%) changes in grain growth rate \cite{kelly2018anti}. Another investigation of GB mobility anisotropy pointed out the GB mobility anisotropy transition temperature lies in the abnormal grain growth regime and can impact the fast-type and low-type GBs. However, it does not seem as the grain growth transition is caused by a change in the GB anisotropy. Instead, studies suggest that the transition is associated with a change in the GB stoichiometry, while the atomistic GB structure remains largely unchanged through the grain growth transition \cite{sternlicht2016mechanism,shih2010investigation, baurer2010abnormal,sternlicht2019characterization,sternlicht2019characterization2}.
Additionally, very recent experimental observations demonstrate that the abnormal grain growth in Fe-doped STO during sintering is strongly influenced by dopant concentration \cite{zahler2023grain}. Nominally undoped STO, exhibits a monomodal grain size distribution despite the presence of GB misorientation. Hence, this concentration-dependent grain size distribution highlights the critical role of solute drag in driving abnormal grain growth, rather than GB misorientation.}

{Accordingly, we develop a phase-field grain growth model grounded in the principles of defect chemistry to reproduce SCL formation and investigate the role of solute drag effects during the sintering process. We exclude additional influences, such as GB misorientation on the grain growth process, on purpose, and in this way, we demonstrate that, even in the absence of these factors, the essential features of {skewed grain size distribution which do not follow the normal grain growth} can still be observed solely through the solute drag effect. The novelty of this work lies in three key aspects. First, this phase-field model accounts for unique segregation energies and distinct site densities of different defects in both the bulk phase and GB core, ensuring that the defect chemistry of both bulk and core are treated in a thermodynamically consistent description. Second, the capabilities of the phase-field model are rigorously benchmarked against sharp interface calculations, providing validation and accuracy. Third, the phase-field results reveal that solute drag alone is sufficient to lead to {the skewed grain size distribution}, confirming its role as the essential mechanism underlying this phenomenon.}

The outline of this paper is as follows. \Cref{defect chemistry} presents the fundamentals of defect chemistry and the classical sharp interface bicrystal models, i.e., the GC and MS models, for they serve as important benchmarks in later sections. \Cref{PF model} describes the grain growth phase-field model in the principle of defect chemistry and electrostatics. Then, we demonstrate the implementation of the phase-field model via the finite element method in \Cref{FEM phase field}. The capabilities of this model are discussed for bicrystal and polycrystalline cases in \Cref{FEM_phasefield_simulations}. Thereby the solute drag induced {skewed grain size distribution} are systematically presented and discussed, particularly in terms of the impact of different defect chemistry parameters. 
It is followed by a concluding \Cref{Conclusion}.

\section{Defect {chemistry} of oxide ceramics}\label{defect chemistry}

{In this section, we review the classical defect chemistry theory in determining the SCL formation under equilibrium state. The performed defect chemistry calculations are used to benchmark the phase-field simulations.} The basics of defect chemistry and the sharp interface bicrystal models (the MS and the GC models) for the equilibrium state were well described in Refs. \cite{de2009formation, tong2020analyzing, maier2023physical}. 
{According to abrupt GB$|$SCL model of Jamnik et al. \cite{jamnik1995interfaces}, the interface region is divided into a GB core and SCL region, in which the materials constants behave as step functions.
The two phases (GB core and bulk) are essentially treated as two phases, each with their own properties, and separated by a sharp interface.} Thus the concentration profiles of oxygen vacancy and dopant , for $\mu^0\V \neq 0$ and $\mu^0\dop \neq 0$, are discontinuous, with concentration jumps across the GB core. For the bicrystal, it becomes a one-dimensional case. 

Assume a {symmetric bicrystal} in a coordinate system $x$, with the origin $x=0$ located at the GB core. {We introduce $c_\text{def}$ point defects into $N_\text{def}$ (the subscript def denotes different types of point defects) available number sites in a unit volume. We should note that $N_\text{def}$ varies for different sublattices and different phases. 
When $N_\text{def} \gg c_\text{def}$, point defects are assumed to be dilute and non-interacting {in both GB core and bulk phases}. }
{In Ref. \cite{maier2023physical}, the electrochemical potential of defect is formulated as
\begin{equation}\label{dc_ep}
    \mu_\text{def} = \mu^0_\text{def} + k_\text{B} T \ln\frac{c_\text{def}}{N_\text{def} - c_\text{def}} + z_\text{def}e\phi,
\end{equation}
where $\mu^0_\text{def}$ is the standard formation energy of defect. $k_\text{B}$, $T$, $z_\text{def}$, $e$ and $\phi$ are the Boltzmann constant, temperature, valance state of the defect, the elementary charge and the electrostatic potential, respectively. }
Then, based on the abrupt GB$|$SCL model aforementioned, the electrochemical potentials of oxygen vacancy and dopant in the bulk and in the core region can be formulated \cite{de2009formation} as
\begin{equation} \label{mu_V_b_x}
\mu\Vb(x) = \mu\Vb^0 + k_\text{B} T \ln \left[\frac{c\Vb(x)}{N\Vb-c\Vb(x)}\right] + z\V e\phi(x),
\end{equation}

\begin{equation}
\mu\dopb(x) = \mu\dopb^0 + k_\text{B} T \ln \left[\frac {c\dopb(x)}{N\dopb-c\dopb(x)}\right] + z\dop e\phi(x),
\end{equation}

\begin{equation}
\mu\Vc = \mu\Vc^0 + k_\text{B} T \ln \left[\frac{c\Vc}{N\Vc-c\Vc}\right] + z\V e\Phi_0,
\end{equation}

\begin{equation}\label{mu_dopc}
\mu\dopc = \mu\dopc^0 + k_\text{B} T \ln \left[\frac{c\dopc}{N\dopc-c\dopc}\right] + z\dop e\Phi_0,
\end{equation}
 where $\mu\Vb^0$,  $\mu\Vc^0$, $\mu\dopb^0$ and $\mu\dopc^0$ denote the standard formation energies of oxygen vacancy and dopant in the bulk and in the core, respectively. The symbols $N\Vb^0$,  $N\Vc^0$, $N\dopb^0$ and $N\dopc^0$ indicate the number of available sites per unit volume of oxygen vacancy and dopant in the bulk and core, respectively. $z\V$ and $z\dop$ denote the charge numbers of oxygen vacancy and acceptor dopant. {$\phi(x)$ is the spatially various electrostatic potential in the bulk phase, and $\Phi_0$ is the electrostatic potential in the GB core, which taken to be constant.} Similarly, $c\dopc$ and $c\Vc$ are the concentrations of dopant and oxygen vacancy in the GB core. Thus, we have $\Phi_0 = \phi(0)$, $c\dopc = c\dopb(0)$ and $c\Vc = c\Vb(0)$.  The formation of a SCL requires the segregation of oxygen vacancy in the core. The smaller value of $\mu\Vc^0$ than $\mu\Vb^0$, i.e. $\mu\Vc^0-\mu\Vb^0<0$, drives the formation of SCL. At the equilibrium state, when electrochemical potentials should be everywhere the same (in other words, no gradient), we have
 \begin{equation}
\mu\Vc = \mu\Vb(x) = \mu\Vb(\infty),
 \end{equation}
  \begin{equation}
\mu\dopc = \mu\dopb(x) = \mu\dopb(\infty),
 \end{equation}
Hence, equilibrium concentration profiles of oxygen vacancy and dopant in the SCL take the form 
\begin{equation}\label{c_V_b}
c\Vb(x) = c\Vb(\infty)\exp\left[\frac{-z\V e\{\phi(x)-\phi(\infty)\}}{k_\text{B} T}\right],
\end{equation}
\begin{equation}\label{c_dop_b}
c\dopb(x) = \frac{c\dopb(\infty)N\dopb\exp\left[\frac{-z\dop e\{\phi(x)-\phi(\infty)\}}{k_\text{B} T}\right]}{N\dopb + c\dopb(\infty)\left\{\exp\left[\frac{-z\dop e\{\phi(x)-\phi(\infty)\}}{k_\text{B} T}\right] -1\right\}}.
\end{equation}
In analogy, equilibrium concentrations of oxygen vacancy and dopant in the core are 
\begin{equation} \label{c_V_c}
c\Vc = \frac{N\Vc c\Vb(\infty) \exp\left[-\frac{\Delta g\V + z\V e\Phi_0}{k_\text{B} T}\right]}{N\Vb + c\Vb(\infty)\exp\left[-\frac{\Delta g\V+ z\V e\Phi_0}{k_\text{B} T}\right]},
\end{equation}
\begin{equation} \label{c_dop_c}
c\dopc = \frac{N\dopc c\dopb(\infty) \exp\left[-\frac{\Delta g\dop+ z\dop e\Phi_0}{k_\text{B} T}\right]}{N\dopb + c\dopb(\infty)\exp\left[-\frac{\Delta g\dop+ z\dop e\Phi_0}{k_\text{B} T}\right]},
\end{equation}
 where $\Delta g\V = \mu\Vc^0 - \mu\Vb^0$ and $\Delta g\dop = \mu\dopc^0 - \mu\dopb^0$. All parameters are tabulated in  \Cref{shape interface parameters}.

\begin{table}[htb]   
\begin{center}   
\caption{Summary of the parameters in the sharp interface method}  
\label{shape interface parameters}
\begin{tabular}{l l c}   
\hline   \textbf{Parameter} & \textbf{Definition} & \textbf{Unit} \\  
\hline   $\mu\Vb^0$ & Formation energy of oxygen vacancy in the bulk phase & eV  \\  
  $\mu\Vc^0$ & Formation energy of oxygen vacancy in the core region& eV\\
   $\mu\dopb^0$ & Formation energy of dopant in the bulk phase  & eV \\
   $\mu\dopc^0$ & Formation energy of dopant in the core region & eV \\
   $\Delta g\V$ & \makecell[l l] {The difference of formation energy of oxygen vacancy between core and bulk,\\ i.e. $\Delta g\V = \mu\Vc^0 - \mu\Vb^0$ }& eV \\
   $\Delta g\dop$ & \makecell[l l] {The difference of formation energy of dopant between core and bulk, \\ i.e. $\Delta g\dop = \mu\dopc^0 - \mu\dopb^0$} & eV \\
   $N\Vb$ &  Number of available sites of oxygen vacancy in the bulk phase & $\frac{1}{\text{m}^3}$ \\
   $N\Vc$ & Number of available sites of oxygen vacancy in the core region & $\frac{1}{\text{m}^3}$ \\
   $N\dopb$ & \makecell[l]{Number of available sites of dopant in the bulk phase.\\ $N\dopb = 1/a^3$, with $a$ being the perovskite lattice constant} & $\frac{1}{\text{m}^3}$ \\
   $N\dopc$ & Number of available sites of dopant in the core region & $\frac{1}{\text{m}^3}$ \\
   $z\V$ & Valency number of oxygen vacancy, $z\V=2$ & - \\
   $z\dop$ & Valency number of dopant, $z\dop=-1$ for $\text{Fe}_\text{Ti}^{'}$ in Fe-doped STO & -\\
   $c\dopb(\infty)$&Number density of dopant in the bulk far from the core region& $\frac{1}{\text{m}^3}$ \\
   $c\Vb(\infty)$ & Number density of oxygen vacancy in the bulk far from the core region & $\frac{1}{\text{m}^3}$ \\
   $k_\text{B}$ & Boltzmann constant $k_\text{B} = 1.38\times 10^{-23}$ & $\frac{\text{J}}{\text{K}}$\\
   $T$ & Temperature & K \\
   $\epsilon_0$ & Vacuum permittivity $\epsilon_0=8.854\times 10^{-12}$ & $\frac{\text{C}}{\text{Vm}}$\\
   $\epsilon_\text{r}$ &  Relative permittivity&  - \\
   $w_\text{c}$ & width of the core & m \\
   $l_\text{D}$ & Debye length & m \\
\hline   
\end{tabular} 
\end{center}   
\end{table}

\subsection{The Mott-Schottky model}
The Mott-Schottky model assumes that the dopant is immobile and its concentration remains fixed and constant everywhere, i.e. $\nabla c\dop = 0$. Note that this assumption simplifies the scenario but breaks the equilibrium condition of the electrochemical potential of the dopants. Based on this assumption, Poisson's equation can be written as
\begin{equation} \label{MS equation}
\begin{split}
\epsilon_0\epsilon_\text{r}\frac{\partial^2\phi}{\partial x^2} &=-\rho(x)\\
& = -e[z\dop c\dopb (\infty) + z\V c\Vb(x)].
\end{split}
\end{equation}
Substituting $c\dopb(\infty) = 2 c\Vb(\infty)$ into Eq. $\eqref{MS equation}$, we have
\begin{equation} \label{MS2}
\epsilon_0\epsilon_\text{r}\frac{\partial^2\phi}{\partial x^2} = 2ec\Vb(\infty)\left\{1-\exp\left[\frac{-2e\{\phi(x)-\phi(\infty)\}}{k_\text{B} T}\right]\right\}.
\end{equation}
With $l_\text{D} = \sqrt{\frac{\epsilon_0\epsilon_\text{r} k_\text{B} T}{2e^2c\dopb(\infty)}}$ being the Debye length and $\alpha =\frac{e}{k_\text{B} T}$, Eq.$\eqref{MS2}$ becomes
\begin{equation} \label{MS3}
\frac{\partial^2 \phi}{\partial x^2} = \frac{1}{2\alpha l_\text{D}^2}\left\{1-\exp\left[\frac{-2e\{\phi(x)-\phi(\infty)\}}{k_\text{B} T}\right] \right\},
\end{equation}

\subsection{The Gouy-Chapman model}
In the Gouy-Chapman model both point defects (oxygen vacancies and dopant) are mobile. The electrochemical potential of the {acceptor} dopant is also considered to be {constant}, as is that of the oxygen vacancies. Thus, Poisson's equation becomes
\begin{equation} \label{GC equation1}
\begin{split}
\epsilon_0\epsilon_\text{r}\frac{\partial^2\phi}{\partial x^2} &=-\rho(x)\\
& = -e[z\dop c\dopb (x) + z\V c\Vb(x)].
\end{split}
\end{equation}
Substituting Eq. \eqref{c_V_b} and \eqref{c_dop_b} into \eqref{GC equation1}, we have
\begin{equation} \label{GC equation2}
\begin{split}
\epsilon_0\epsilon_\text{r}\frac{\partial^2\phi}{\partial x^2} &= -e\left\{z\dop\frac{c\dopb(\infty)N\dopb\exp\left[\frac{-z\dop e\{\phi(x)-\phi(\infty)\}}{k_\text{B} T}\right]}{N\dopb + c\dopb(\infty)\left\{\exp\left[\frac{-z\dop e\{\phi(x)-\phi(\infty)\}}{k_\text{B} T}\right] -1\right\}} \right. \\ 
& \left. + z\V c\Vb(\infty)\exp\left[\frac{-z\V e\{\phi(x)-\phi(\infty)\}}{k_\text{B} T}\right]\right\}.
\end{split}
\end{equation}

The numerical solutions of the MS and GC models have been achieved through an iterative process \cite{de2009formation}. In the present work, we propose an alternative numerical method via finite element method (FEM). By considering the symmetric configuration of a bicrysal (ignoring the grain misorientation), the semi-infinite simulation domain of $[0, \infty)$ is sufficient. The location of the GB core is located at the left boundary of the semi-infinite domain ($x=0$). As for the boundary conditions, the electrostatic field is grounded at the right side, i.e., $\phi|_{x=\infty} = 0$,  while at the left side, a mixed boundary condition is given by
\begin{equation} \label{BC_left}
    \frac{\partial \phi}{\partial x}\bigg|_{x=0} = -\frac{Q_\text{c}(\phi_0)}{2\epsilon_0\epsilon_\text{r}},
\end{equation}
where $Q_\text{c}(\phi_0) = e w_\text{c}(z\V c\Vc + z\dop c\dopc)$ and $\phi_0 = \phi|_{x=0}$. We employed the boundary element in the finite element method to regard directly the mixed type of boundary condition defined in the above equation. The obtained numerical solutions agree fully with these from Ref. \cite{de2009formation} (See Supplementary). 

\section{Phase-field grain growth model with defect chemistry and electrostatics}\label{PF model}
\subsection{Free energy density functional and electrochemical potentials} \label{free energy density functional}

{In this part, we aim to develop a defect-chemistry informed phase-field model including not only the dominant driving force of SCL formation, i.e., segregation energy of oxygen vacancy, but also the segregation energy of acceptor dopant. The segregation energy of acceptor dopant can not govern the SCL formation, whereas it plays an important role on GB potential \cite{usler2024point}, which is barely investigated by phase-field simulations reported in previous literature. In addition, we also include different available number sites of oxygen vacancy and acceptor dopant in bulk and GB core, i.e. $N\Vb \neq N\dopb$ and $N\Vc \neq N\dopc$.}

Assuming an Fe-doped STO polycrystalline system consisting of $n$ grains, we choose {free energy based} Kim-Kim-Suzuki (KKS) phase-field model and utilize a set of non-conserved order parameters (OPs) ${\eta_i}$ to distinguish between different grains \cite{kim1999phase}. 
Then, by treating the GB core and bulk as distinct phases with separate segregation energies \cite{cha2002phase} , the phase-field model presented here aligns with the assumptions of the abrupt GB$|$SCL model (see \Cref{defect chemistry}) and inherently incorporates solute drag effects. Therefore, within grain $i$, we have $\eta_i = 1$ and $\eta_j = 0$ for $j < n$ and $j \neq i$. Within the GB core region of grain $i$ and $j$, the order parameters $\eta_i$ and $\eta_j$ spatially vary from 0 to 1. {In the latter context, $\eta$ denotes the entire set of OPs \{$\eta_i$\}.}
When considering electrochemical contributions, we employ two conserved concentration fields along with an electrostatic potential field. Other normalized field variables are $C\V$ for oxygen vacancy concentration, $C\dop$ for dopant concentration, and $\phi$ for the electrostatic potential. 
{In Fe-doped STO, $\text{Fe}^{3+}$ and $\text{Fe}^{4+}$ can coexist. At high temperatures and low oxygen partial pressures, a higher proportion of $\text{Fe}^{3+}$ is observed and thus the dopant $\text{Fe}_\text{Ti}^{'}$ has negative charge. For example, during the sintering process at 1623 K in air, approximately 90\% of the iron is estimated to be in the $\text{Fe}^{3+}$ state, according to the bulk defect chemical model (see Fig. A1 in Ref. \cite{merkle2020water}). Thus, in this study, we assume the valence state of Fe to be 3+  and therefore the valance of dopant $\text{Fe}_\text{Ti}^{'}$ is -1, i.e. $z\dop = -1$.}
Then, the formulation of total free energy density functional is given by
\begin{equation} \label{Free_energy_density_functional}
    \mathscr{F}  =  \int_\Omega (f^\text{grad} + f^\text{loc} + f^\text{ech}) \text{d}\Omega.
\end{equation}
{The free energy density functional contains three contributions. The first term $f^\text{grad}$ represents the gradient energy of the diffuse interface and is given by}
\begin{equation}
f^\text{grad} = \frac{1}{2}\kappa\sum_{i=1}^n|\nabla\eta_i|^2 ,
\end{equation}
where $\kappa$ is the coefficient of the gradient term. {The second term $f^\text{loc}$ is the multi-well potential and it reaches minimum value in the bulk phase. According to Ref. \cite{moelans2008quantitative, aagesen2024electrochemical}, its formulation is given by}
\begin{equation}
f^\text{loc} = \omega\sum_{i=1}^n \left(\frac{\eta_i^4}{4} - \frac{\eta_i^2}{2} + \gamma \sum_{i=1}^n \sum_{j>i}^n \eta_i^2\eta_j^2 + \frac{1}{4}\right),
\end{equation}
 with $\omega$ being the free energy barrier coefficient, $\gamma = 1.5$ for producing symmetric profile of $\eta$ and isotropic grain growth. 

{The determination of parameters such as $\kappa$ and $\omega$ can be linked to characteristics such as GB core width ($w_\text{c}$) and GB energy ($\sigma$) \cite{moelans2008quantitative}.} Their relations can be expressed as
\begin{equation}
    \sigma = \frac{\sqrt{2}}{3}\sqrt{\kappa\omega},
\end{equation}
and
\begin{equation}
    w_\text{c} = \sqrt{\frac{8\kappa}{\omega}}.
\end{equation}
{ In the present work, we only consider the isotropic GB energy, i.e. it is not dependent on GB misorientation.}
The third term in free energy functional is the electrochemical contribution. Here, we treat the electrochemical free energy density as a mixture of bulk phase and GB core by an interpolation function. Thus, {with the consideration of electrostatic free energy according to Refs. \cite{bishop2003effect, garcia2004thermodynamically},} $f^{\text{ech}}$ is
\begin{equation} \label{f_chem}
f^\text{ech} = [1-h(\eta)] f^\text{ech}_\text{b}(C\Vb, C\dopb)  + h(\eta) f^\text{ech}_\text{c}(C\Vc, C\dopc) - \frac{\epsilon_0\epsilon_\text{r}}{2}(\nabla\phi)^2,
\end{equation}
where the subscripts b and c indicate the bulk phase and GB core, respectively. $h(\eta)$ is the interpolation function and is given by
\begin{equation}
h(\eta) = \frac{4}{3}\left[1-4\sum_{i=1}^n \eta_i^3 + 3\left(\sum_{i=1}^n \eta_i^2\right)^2\right].
\end{equation}
Here, $h(\eta)$ exhibits non-monotonic behavior. Specifically, for an $i/j$ GB, $h(\eta_i) = 0$ when $\eta_i = 1$ (representing the bulk phase), whereas $h(\eta_i) = 1$ when $\eta_i = 0.5$ (representing the core). This interpolation function allows us to differentiate between the bulk phase and the core region. $\epsilon_0$ is the vacuum permittivity; $\epsilon_\text{r}$ is the relative permittivity. {For STO, $\epsilon_\text{r} = 9\times 10^4/(T-35)$ \cite{de2009formation}. $\epsilon_\text{r}$ is only dependent on temperature in the present work. Hence, $\epsilon_\text{r}$ is constant across the SCL zone for specific temperature.}
Furthermore, $f^{\text{ech}}_\text{b}(C\Vb, C\dopb)$ and $f^{\text{ech}}_\text{c}(C\Vc, C\dopc)$ in \Cref{f_chem} denote the electrochemical free energy contributions from the bulk and core, respectively, with $C\Vb$ and $C\dopb$ being the phase concentrations of oxygen vacancy and dopant in the bulk phase, while $C\Vc$ and $C\dopc$ being the phase concentrations of oxygen vacancy and dopant in the core. When $h(\eta) = 1$, $C\dopc$ and $C\Vc$ correspond to $c\dopc$ and $c\Vc$.  When $h(\eta) = 0$, $C\dopb$ and $C\Vb$ are same as $c\dopb(x)$ and $c\Vb(x)$. Local concentration is treated as the mixture of phase concentrations in bulk and core. The relations between local concentration and phase concentration are expressed as
\begin{equation}\label{localCV}
    C\V = [1-h(\eta)]C\Vb +h(\eta) C\Vc,
\end{equation}
\begin{equation}\label{localCdop}
    C\dop = [1-h(\eta)]C\dopb +h(\eta) C\dopc.
\end{equation}
{In principle of defect chemistry thermodynamic model as mentioned in \Cref{defect chemistry}, the  electrochemical free energies $f^{\text{ech}}_\text{c}(C\Vc, C\dopc)$ and $f^{\text{ech}}_\text{b}(C\Vb, C\dopb)$  deliver the required equilibrium thermodynamic information in bulk and GB core and their expressions are given by}
\begin{equation} \label{ech_c}
\begin{split}
f^\text{ech}_\text{c} (C\Vc, C\dopc) &= \frac{1}{V_\text{m}}\left\{g^0\Vc C\Vc + g^0\dopc C\dopc \right. \\
&+ RT \left[C\Vc\ln\left(\frac{C\Vc}{\tilde{N}\Vc}\right) + (\tilde{N}\Vc-C\Vc)\ln\left(\frac{\tilde{N}\Vc-C\Vc}{\tilde{N}\Vc}\right)  \right.\\
 &\left. \left. + C\dopc\ln\left(\frac{C\dopc}{\tilde{N}\dopc}\right) + (\tilde{N}\dopc-C\dopc)\ln\left(\frac{\tilde{N}\dopc-C\dopc}{\tilde{N}\dopc}\right) \right]\right\} \\
 &+ \frac{\mathcal{F}}{V_\text{m}}\left(z\V C\Vc + z\dop C\dopc\right) \phi,
\end{split}
\end{equation}
\begin{equation}\label{ech_b}
\begin{split}
f^\text{ech}_\text{b} (C\Vb, C\dopb) &= \frac{1}{V_\text{m}}\left\{g^0\Vb C\Vb + g^0\dopb C\dopb \right. \\
&+ RT \left[C\Vb\ln\left(\frac{C\Vb}{\tilde{N}\Vb}\right) + (\tilde{N}\Vb-C\Vb)\ln\left(\frac{\tilde{N}\Vb-C\Vb}{\tilde{N}\Vb}\right) \right.\\
 &\left. \left. + C\dopb\ln\left(\frac{C\dopb}{\tilde{N}\dopb}\right) + (\tilde{N}\dopb-C\dopb)\ln\left(\frac{\tilde{N}\dopb-C\dopb}{\tilde{N}\dopb}\right) \right]\right\}\\
 &+ \frac{\mathcal{F}}{V_\text{m}}\left(z\V C\Vb + z\dop C\dopb\right) \phi,
\end{split}
\end{equation}
where $g^0\Vc$, $g^0\Vb$, $g^0\dopc$, $g^0\dopb$ are standard formation energies of oxygen vacancy and dopant in bulk and core.  $\tilde{N}\Vc$, $\tilde{N}\Vb$, $\tilde{N}\dopc$, $\tilde{N}\dopb$ are dimensionless numbers of available sites per unit volume for oxygen vacancy and dopant in bulk and core. $z\V$ and $z\dop$ are valance numbers of oxygen vacancy and dopant.   $\mathcal{F}$, $V_\text{m}$, $R$, $T$ are the Faraday constant, molar volume, gas constant and temperature, respectively. {Based on experimental evidence of abnormal grain growth in 0.2\% Fe-doped STO \cite{zahler2023grain}, where defect-defect interactions are negligible, we limit our consideration to the dilute defect case. This simplification ensures that the fundamental characteristics of the system are captured without introducing unnecessary complexities. However, the dilute assumption is not valid when high defect concentration is involved, the interactions between different defects should be considered \cite{chen2022thermodynamic}.} 

The electrochemical potentials of oxygen vacancy and dopant in bulk and core are obtained according to
\begin{equation} \label{mu_V_b_PF}
    \mu\Vb^{\text{PF}} = \frac{\delta \mathscr{F}}{\delta C\Vb} = \frac{g\Vb^0}{V_\text{m}} + \frac{RT}{V_\text{m}}\ln{\left(\frac{C\Vb}{\tilde{N}\Vb -C\Vb}\right)} + \frac{\mathcal{F}}{V_\text{m}}z\V\phi,
\end{equation}
\begin{equation}\label{mu_V_c_PF}
    \mu\Vc^{\text{PF}} = \frac{\delta \mathscr{F}}{\delta C\Vc} = \frac{g\Vc^0}{V_\text{m}} + \frac{RT}{V_\text{m}}\ln{\left(\frac{C\Vc}{\tilde{N}\Vc-C\Vc}\right)} + \frac{\mathcal{F}}{V_\text{m}}z\V\phi,
\end{equation}
\begin{equation}\label{mu_dop_b_PF}
   \mu\dopb^{\text{PF}} = \frac{\delta \mathscr{F}}{\delta C\dopb} = \frac{g\dopb^0}{V_\text{m}} + \frac{RT}{V_\text{m}}\ln{\left(\frac{C\dopb}{\tilde{N}\dopb-C\dopb}\right)} + \frac{\mathcal{F}}{V_\text{m}}z\dop\phi,
\end{equation}
\begin{equation}\label{mu_dop_c_PF}
    \mu\dopc^{\text{PF}} = \frac{\delta \mathscr{F}}{\delta C\dopc} = \frac{g\dopc^0}{V_\text{m}} + \frac{RT}{V_\text{m}}\ln{\left(\frac{C\dopc}{\tilde{N}\dopc-C\dopc}\right)} + \frac{\mathcal{F}}{V_\text{m}}z\dop\phi,
\end{equation}
which are consistent with the standard formations of electrochemical potentials in defect chemistry, {i.e. \Cref{mu_V_b_x} to \Cref{mu_dopc}}.

\subsection{Governing equations}
{As mentioned in the introduction, we focus on the solute drag effect, which is the most fundamental and primary factor influencing grain growth patterns. Therefore, our phase-field model employs constant GB energy and mobility and uses order parameters to distinguish different grains without explicitly considering grain misorientation. Then, the evolution of phase-field OPs for grain $i$ with isotropic GB properties becomes}
\begin{equation} \label{detadt}
    \frac{1}{L}\frac{\partial \eta_i}{\partial t} = \frac{\delta \mathscr{F}}{\delta \eta_i},
\end{equation}
where $L$ is {the GB mobility}. Substituting Eq. \eqref{Free_energy_density_functional} into \eqref{detadt}, we can obtain
\begin{equation}\label{detadt1}
\begin{split}
    \frac{1}{L}\frac{\partial \eta_i}{\partial t} &= \nabla\cdot\kappa\nabla\eta_i - \omega \frac{\partial f^{\text{loc}}(\eta)}{\partial \eta_i} \\
    &+ \frac{\partial h(\eta)}{\partial \eta_i}\left[ f_\text{b}^{\text{ech}} - f_{\text{c}}^{\text{ech}} - \frac{\partial f_\text{b}^{\text{ech}}}{\partial C\Vb} (C\Vb - C\Vc)-  \frac{\partial f_\text{b}^{\text{ech}}}{\partial C\dopb} (C\dopb - C\dopc) \right].
\end{split}
\end{equation}
At the quasi-equilibrium state, the core moves with a constant velocity, $v_c$, we have $\dot{\eta} = -v_c\eta'(x)$, thus Eq. \eqref{detadt} becomes
\begin{equation}
\begin{split}
    \frac{v_c}{L} \frac{d\eta}{dx} &= \nabla\cdot\kappa\nabla\eta_i - \omega \frac{\partial f^{\text{loc}}(\eta)}{\partial \eta_i} \\
    &+ \frac{\partial h(\eta)}{\partial \eta_i}\left[ f_\text{b}^{\text{ech}} - f_{\text{c}}^{\text{ech}} - \frac{\partial f_\text{b}^{\text{ech}}}{\partial C\Vb} (C\Vb - C\Vc)-  \frac{\partial f_\text{b}^{\text{ech}}}{\partial C\dopb} (C\dopb - C\dopc) \right].
\end{split}
\end{equation}
{By combining Fick's first and second laws, the evolution of the conserved concentration fields $C\V$ and $C\dop$ are expressed as}
\begin{equation} \label{dcvdt}
    \frac{\partial C\V}{\partial t} = \nabla \left(M\V \nabla\frac{\delta \mathscr{F}}{\delta C\V}\right), 
\end{equation}
\begin{equation} \label{dcdopdt}
    \frac{\partial C\dop}{\partial t} = \nabla \left(M\dop \nabla\frac{\delta \mathscr{F}}{\delta C\dop}\right), 
\end{equation}
where $M\V$ and $M\dop$ are mobilities of oxygen vacancy and dopant, which are given by $M\V= \frac{D\V}{\partial^2 f^{\text{ech}}/\partial C\V^2}$ and $M\dop= \frac{D\dop}{\partial^2 f^{\text{ech}}/\partial C\dop^2}$, with $D\V$ and $D\dop$ being the diffusivities of oxygen vacancy and dopant, respectively. 
Moreover, local equilibrium between bulk phases and core should be satisfied everywhere. Therefore, we have following constrains
\begin{equation}\label{eqV}
   \delta \mathscr{F}/\delta C\V = \delta \mathscr{F}/\delta C\Vb = \delta \mathscr{F}/\delta C\dopb,
\end{equation}
\begin{equation}\label{eqdop}
\delta \mathscr{F}/\delta C\dop = 
    \delta \mathscr{F}/\delta C\Vc = \delta \mathscr{F}/\delta C\dopc.
\end{equation}
Therefore, $\delta \mathscr{F}/\delta C\V$ and $\delta \mathscr{F}/\delta C\dop$ in \Cref{dcvdt,dcdopdt} can be substituted by $\mu\Vb^{\text{PF}}$, $\mu\dopb^{\text{PF}}$ or $\mu\Vc^{\text{PF}}$, $\mu\dopc^{\text{PF}}$.
In addition, the governing equation of electrostatic potential field is obtained by Poisson's equation
{
\begin{equation} \label{PoissonEquation}
\begin{split}
\epsilon_0\epsilon_\text{r} \nabla^2 \phi &= - \frac{\mathcal{F}}{V_\text{m}}(z\V C\V + z\dop C\dop )\\
 &= - \frac{\mathcal{F}}{V_\text{m}}\left\{z\V C\Vb[1-h(\eta)] + z\V C\Vc h(\eta)  + z\dop  C\dopb[1-h(\eta)] + z\dop C\dopc h(\eta). \right\}\\
\end{split}
\end{equation}
Substituting the site fractions $[V\Vb]$, $[V\Vc]$, $[V\dopb]$ and $[V\dopc]$ into \Cref{PoissonEquation}, we have
\begin{equation}\label{Poission2}
\begin{split} 
    \epsilon_0\epsilon_\text{r} \nabla^2 \phi = & - \frac{\mathcal{F}}{V_\text{m}}\left\{z\V \tilde{N}\Vb[V\Vb][1-h(\eta)] + z\V \tilde{N}\Vc[V\Vc] h(\eta)  \right. \\
 & \left. + z\dop  \tilde{N}\dopb[V\dopb][1-h(\eta)] + z\dop \tilde{N}\dopc[V\dopc] h(\eta)  \right\}.
 \end{split}
 \end{equation}
 We clearly observe the numbers of available sites $\tilde{N}\Vb$, $\tilde{N}\Vc$, $\tilde{N}\dopb$, $\tilde{N}\dopc$ influence the electrostatic potential distribution in \Cref{Poission2}. 
 The parameters utilized in phase-field model are tabulated in \cref{PF_paramters}.

{
\subsection{Comparison to other grain growth phase-field models for oxide ceramics}} 
 \label{PF_models_Comparison}
{In this part, we outline differences between our defect-chemistry-informed phase-field model, the model developed by Vikrant et al. \cite{vikrant2020electrochemical} and the grand-potential based phase-field model developed by Aagesen et al. \cite{aagesen2024electrochemical}.}

{
In accordance with defect chemistry theory \cite{maier2023physical}, we include the available site densities for the bulk and the grain boundary core explicitly, as shown by the bulk and GB core electrochemical free energy (\Cref{ech_c} and \Cref{ech_b}) and the oxygen vacancy and dopant electrochemical potentials in bulk and GB core (\Cref{mu_V_b_PF}-\Cref{mu_dop_c_PF}). 
It is important to do so, because they can impact the space charge layers and GB potential considerably \cite{usler2024point}. As mentioned in the Introduction, in Fe-doped SrTiO$_3$, Fe substitutes for Ti$^{4+}$ on the B-site, and charge compensation is typically provided by oxygen vacancies on the O-site. This results in the number density of available sites for the dopant in the bulk that is one-third of that for  an oxygen vacancy. Even more important, B-site dopant and O-site oxygen vacancy belong to different sub-lattices and thus have to be accounted for separately, in order to correctly represent the respective configurational entropy. In addition, the number of available sites for both species will deviate significantly from their bulk values due to the distinct atomic structure of the GB core. Therefore, it is necessary to treat the available site densities for oxygen vacancies and dopants in the GB core as distinct from those in the bulk, and these quantities play a critical role in determining the space charge layer. As our simulation results in the following sections demonstrate, the different numbers of available site significantly influence both the grain boundary potential and the defect concentration distributions, consequently, leading to different space charge layers formation. Much more results can be found in Figures 2 and 3, validated against De Souza’s work \cite{de2009formation}. }

{
Such explicit incorporation of the numbers of available site of oxygen vacancy and dopant in the bulk and GB core is not fully regarded in the existing models. For instance, Vikrant et al. assumed that the oxygen vacancy and acceptor dopant share the same sublattice in the chemical free energy contribution (see Eq. 1 in Ref. \cite{vikrant2020electrochemical}).
\begin{equation} \label{Vikrant_free_energy}
\begin{split}
f^\text{Vik}(\eta, [V\V], [V\dop], T) &= \frac{1}{\omega} \left\{
f\V(\eta, T)[V\V] + f_{\text{dop}}(\eta, T)[V\dop] + k_\text{B} T \left( [V\V]\ln[V\V] + [V\dop]\ln[V\dop] \right) \right. \\
& \left. + k_\text{B} T \left( 1 - [V\V] - [V\dop] \right) \ln\left( 1 - [V\V] - [V\dop] \right)  + \Omega_{\text{V}_{\text{O}}^{..} \text{dop}}[V\V][V\dop]\right\} ,
\end{split}
\end{equation}
where $f\V(\eta, T)$ and $f\dop(\eta, T)$ are temperature related formation energy of oxygen vacancy and dopant in bulk and GB core. $[V\V]$ and $[V\dop]$ are the site fractions of oxygen vacancy and dopant. $\Omega_{\text{V}_{\text{O}}^{..} \text{dop}}$ is the interaction coefficient between oxygen vacancy and dopant. At here, we omit  $\Omega_{\text{V}_{\text{O}}^{..} \text{dop}}$ for dilute case. In \Cref{Vikrant_free_energy}, both oxygen vacancy and dopant are described as occupying the same sublattice, which is reflected in the entropic mixing term $\left( 1 - [V\V] - [V\dop] \right) \allowbreak \ln\left( 1 - [V\V] - [V\dop] \right)$.
Then, we can obtain the electrochemical potential of oxygen vacancy and dopant from Vikrant's model
\begin{equation}
\mu\V^\text{Vik} = \frac{f\V(\eta, T)}{\omega} + \frac{k_\text{B}T}{{\omega}}\ln \left(\frac{[V\V]}{1-[V\V] - [V\dop]}\right) +  z\V e \phi,
\end{equation}
\begin{equation}
    \mu\dop^\text{Vik} = \frac{f\dop(\eta, T)}{\omega} + \frac{k_\text{B}T}{\omega}\ln \left(\frac{[V\dop]}{1-[V\V] - [V\dop]}\right) +  z\dop e \phi ,
\end{equation}
which is different from the electrochemical potential obtained from defect-chemistry as shown in \Cref{dc_ep}-\Cref{mu_dopc} and present work in \Cref{mu_V_b_PF} - \Cref{mu_dop_c_PF}. In contrast, our model explicitly distinguishes between the numbers of available site of oxygen vacancy and dopant in both bulk and GB core, i.e. $\tilde{N}\Vb$, $\tilde{N}\Vc$, $\tilde{N}\dopb$, $\tilde{N}\dopc$. The configurational entropy term in the electrochemical free energy part is $\left(\tilde{N}_i - C_i\right)\ln\left(\frac{\tilde{N}_i-C_i}{\tilde{N}_i}\right)$, with the subscript $i$ indicates $\text{V}_\text{O}^{..},\text{b}$, $\text{V}_\text{O}^{..},\text{c}$, $\text{dop},\text{b}$ and $\text{dop},\text{c}$. This separation allows us to represent the thermodynamics of defect segregation more accurately at grain boundary core. 
Moreover, the numbers of available site of oxygen vacancy and acceptor dopant in both the bulk and GB core explicitly influence the Poisson's equation and the electrostatic potential distribution in our model (see \Cref{Poission2}), However, in Vikrant's model, the Poisson's equation is $\epsilon_0\epsilon_r\nabla^2\phi = -e(z\V[V\V] + z\dop[V\dop])$, the influence of these parameters, $\tilde{N}\Vb$, $\tilde{N}\Vc$, $\tilde{N}\dopb$, $\tilde{N}\dopc$, on the electrostatic potential distribution is not considered.}

{Aagsen's model is developed based on the grand potential functional. The thermodynamic equivalence between the free energy based KKS model and grand potential based phase-field model developed has been clarified in Ref. \cite{plapp2011unified} (see section \textbf{V}A). However, in Aagsen's model, the numbers of available site (denoted as $n_\text{Y}$ and $n_\text{O}$) of both cation vacancy and oxygen vacancy  in Y$_2$O$_3$ are treated as uniform across the bulk and GB core, i.e. $n_\text{Y,b} = n_\text{Y,c}$ and $n_\text{O,b} = n_\text{O,c}$, which may overlook the important structural variations at GB core, especially when $n_\text{Y,b} \neq n_\text{Y,c}$ and $n_\text{O,b} \neq n_\text{O,c}$, and result in inappropriate space charge layer formation. These scenarios have been comprehensively discussed in De Souza's work \cite{de2009formation} and demonstrated in the present work (see \Cref{fig3_new} and \Cref{fig5_new}) in \Cref{equilibrium}.}

{In addition, such explicit consideration of numbers of available site in our formulation paves the foundation for the subsequent extension towards the treatment of charge state transition of dopants, such as Fe$^{4+}$/Fe$^{3+}$ and Fe$^{3+}$/Fe$^{2+}$ transitions in Fe-doped SrTiO$_3$ \cite{klein2023fermi,bonkowski2024single}. Vikrant's work, Aagsen's work and the present work primarily focus on a single valence state (typically -1), which limits their applicability under varying thermodynamic conditions. The dominant charge state of Fe can vary significantly with temperature and oxygen partial pressure \cite{wang2016defect}. These charge states all share the same B-site sublattice in the perovskite structure, which reinforces the importance of correctly defining and distinguishing the numbers of available site between oxygen vacancy and dopant in bulk phase and grain boundary core in the phase-field model. Without completely considering the defect chemistry, the extension of Vikrant's model or Aagsen's model in this direction is not applicable due to the mixture of the numbers of available site. However, our defect-chemistry phase-field is developed to accommodate this level of resolution, enabling us to simulate space charge layer formation with proper consideration of the different valence states of dopant and their impact on different space charge layer formations.}


\begin{table}[p]   
\begin{center}   
\caption{Summary of the parameters in the phase-field model}  
\label{PF_paramters} 
\begin{tabular}{l l c}   
\hline   \textbf{DOFs} & \textbf{Definition} & \textbf{Unit} \\
\hline      $\eta_i$   &   phase-field order parameter of grain $i$, $i=1,2,3,...,n$   &  -      \\  
  $C\V$   &\makecell[l]{ Normalized local concentration of oxygen vacancy  \\($C\V= a^3c\V$, with $a$ being the lattice constant)}&   -   \\
   $C\dop$   &  Normalized local concentration of dopant, $C\dop= a^3c\dop $  & - \\
    $C\Vb$   & Normalized concentration of oxygen vacancy in bulk phase, $C\Vb= a^3c\Vb $& - \\
    $C\Vc$   & Normalized concentration of oxygen vacancy in core region, $C\Vc= a^3c\Vc $ & - \\
    $C\dopb$   & Normalized concentration of dopant in bulk phase, $C\dopb= a^3c\dopb$ & - \\
    $C\dopc$   & Normalized concentration of dopant in core region, $C\dopc= a^3c\dopc$ & - \\
   ${\phi}$   & Electrostatic potential & V \\
\hline   
\hline   \textbf{Parameters} & \textbf{Definition} & \textbf{Unit} \\
\hline  $\kappa$ & Coefficient of gradient term & $\frac{\text{J}}{\text{m}} $ \\
$\omega$& Heigh of well potential  & $\frac{\text{J}}{\text{m}^3}$ \\
$L$ & GB mobility  & $\frac{\text{m}^3}{\text{Js}}$ \\
$g^0\Vb$&\makecell[l]{Formation energy of oxygen vacancy in the bulk phase.\\ $g^0\Vb = \mu\Vb^0 N_\text{A}$, with $N_\text{A}$ being Avogadro constant} & $\frac{\text{J}}{\text{mol}}$\\
$g\Vc^0$& \makecell[l]{Formation energy of oxygen vacancy in the core region.\\ $g^0\Vc = \mu\Vc^0 N_\text{A}$ } & $\frac{\text{J}}{\text{mol}}$\\
$g^0\dopb$&\makecell[l]{Formation energy of dopant in the bulk phase. \\$g^0\dopb = \mu\dopb^0 N_\text{A}$} &   $\frac{\text{J}}{\text{mol}}$\\
$g^0\dopc$&\makecell[l]{Formation energy of dopant in the core region.\\ $g^0\dopc = \mu\dopc^0 N_\text{A}$ }&   $\frac{\text{J}}{\text{mol}}$\\
$V_\text{m}$&Molar volume& $\frac{\text{m}^3}{\text{mol}}$\\
$\tilde{N}\Vc$ &\makecell[l]{ Dimensionless available sites of oxygen vacancy in core region.\\ $\tilde{N}\Vc = N\Vc a^3$, with $a$ being the lattice constant} & - \\
$\tilde{N}\Vb$ & \makecell[l]{ Dimensionless available sites of oxygen vacancy in bulk phase. \\ $\tilde{N}\Vb = N\Vb a^3$} &  - \\
$\tilde{N}\dopc$ & \makecell[l]{Dimensionless available sites of dopant in core region.\\ $\tilde{N}\dopc = N\dopc a^3$ }&-\\
$\tilde{N}\dopb$ & \makecell[l]{Dimensionless available sites of dopant in bulk phase.\\ $\tilde{N}\dopb = N\dopb a^3$ }& - \\
 $R$ & Gas constant & $\frac{\text{J}}{\text{K}}$\\
 $T$ & Temperature & K \\
 $\mathcal{F}$ &Faraday constant & $\frac{\text{C}}{\text{mol}}$ \\
 $D\V$ & Diffusivity of oxygen vacancy & $\frac{\text{m}^2}{\text{s}}$\\
$D\dop$ & Diffusivity of dopant & $\frac{\text{m}^2}{\text{s}}$ \\
\hline
\end{tabular}  
\end{center}   
\end{table}


\section{Finite element implementation of the phase-field model}\label{FEM phase field}

In this section, we present the detailed finite element implementation for the defect-chemistry informed phase-field model. According to the model presented in \Cref{PF model} and \Cref{PF_paramters}, we treat the OPs $\eta_i$ and the field variables $C\V$, $C\dop$, $C\Vb$, $C\Vc$, $C\dopb$, $C\dopc$, $\phi$ as degrees-of-freedom (DOFs). Moreover, 
to solve Eqs. \eqref{dcvdt} and \eqref{dcdopdt}, we use the mixed finite element formulation and treat the electro-chemical potentials $\mu\V$ and $\mu\dop$ as additional DOFs \cite{bai2019two}.

The strong forms of the corresponding governing equations have already been given in Eqs. \eqref{localCV}, \eqref{localCdop} and  \eqref{detadt1} to \eqref{PoissonEquation}. According to the FEM \cite{hughes2012finite}, we formulate the residuals from the weak forms of these governing equations by introducing corresponding {test functions}  $\psi_{C\Vc}$,  $\psi_{C\dopc}$, $\psi_{\eta_i}$, $\psi_{C\V}$, $\psi_{C\dop}$, $\psi_{\mu\V}$, $\psi_{\mu\dop}$, $\psi_{C\Vb}$, $\psi_{C\dopb}$ and $\psi_{\phi}$. 
Firstly, the residuals for the weak forms of Eqs. \eqref{localCV}, \eqref{localCdop} and  \eqref{detadt1} are
\begin{equation}
    R_{C\Vc} = \int_V\psi_{C\Vc}\{C\V -  [1-h(\eta)]C\Vb - h(\eta) C\Vc \}\text{d}V,
\end{equation}

\begin{equation}
    R_{C\dopc} = \int_V\psi_{C\dopc}\{C\dop -  [1-h(\eta)]C\dopb - h(\eta) C\dopc \}\text{d}V,
\end{equation}

\begin{equation}
\begin{split}
    R_{\eta_i} &= \int_V {\psi_{\eta_i}}\frac{\eta_i^{t_{n+1}}-\eta_i^{t_n}}{\Delta t} \text{d}V + L\int_V \psi_{\eta_i} \omega \frac{\partial f^{\text{loc}}(\eta)}{\partial\eta_i} \text{d}V + L\int_V\kappa \eta_{i,j}\psi_{\eta_i ,j}\text{d}V -L\int_\Gamma\kappa\eta_{i,j}\psi_{\eta_i} \hat{n}_j \text{d}\Gamma\\
   &-L\int_V\psi_{\eta_i} \frac{\partial h(\eta)}{\partial \eta_i}\left[ f_\text{b}^{\text{ech}} - f_{\text{c}}^{\text{ech}} - \frac{\partial f_\text{b}^{\text{ech}}}{\partial C\Vb} (C\Vb - C\Vc)-  \frac{\partial f_\text{b}^{\text{ech}}}{\partial C\dopb} (C\dopb - C\dopc) \right]\text{d}V.
    \end{split}
\end{equation}
Note that we use the backward Euler method for the time integration. In the very last equation, $\eta_i^{t_n}$ and $\eta_i^{t_{n+1}}$ indicate $\eta_i$ at the time step $t_n$ and $t_{n+1}$, respectively. Thereby, $\Delta t = t_{n+1} - t_n$, and $\hat{n}_i$ is the normal vector to the boundary $\Gamma$ of the subdomain.
Then, through introducing the addition coupling fields $\mu\V = \delta \mathscr{F}/\delta C\V$ and $\mu\dop = \delta \mathscr{F}/\delta C\dop$, we obtain the weak forms of Eqs. \eqref{dcvdt} and \eqref{dcdopdt} as
\begin{equation}
R_{C\V} = \int_V \psi_{C\V} \left(\mu\V - \frac{\delta  \mathscr{F}}{\delta C\Vb}\right)    \text{d}V.
\end{equation}

\begin{equation}
R_{C\dop} = \int_V \psi_{C\dop} \left(\mu\dop - \frac{\delta  \mathscr{F}}{\delta C\dopb}\right)    \text{d}V.
\end{equation}

\begin{equation}
    R_{\mu\V}=\int_V \psi_{\mu\V} \frac{ C\V^{t_{n+1}} - C\V^{t_n}}{\Delta t} \text{d}V + \int_V  M\V\mu_{V_{O}^{..},i} \psi_{\mu\V, i}\text{d}V - \int_\Gamma  M\V \mu_{V_{O}^{..},i}\psi_{\mu\V} \hat{n}_{i} \text{d}\Gamma.
\end{equation}

\begin{equation}
    R_{\mu\dop}=\int_V \psi_{\mu\dop} \frac{ C\dop^{t_{n+1}}-C\dop^{t_n}}{\Delta t} \text{d}V + \int_V  M\dop\mu_{dop,i} \psi_{\mu\dop, i}\text{d}V - \int_\Gamma  M\dop\mu_{dop,i}\psi_{\mu\dop} \hat{n}_{i} \text{d}\Gamma.
\end{equation}

The weak forms of the local equilibrium constrain in Eqs. \eqref{eqV} and \eqref{eqdop} are  
\begin{equation}
    R_{C\Vb} = \int_V\psi_{C\Vb}\left(\frac{\delta  \mathscr{F}}{\delta C\Vb} - \frac{\delta  \mathscr{F}}{\delta C\Vc}\right) \text{d}V.
\end{equation}

\begin{equation}
    R_{C\dopb} = \int_V\psi_{C\dopb}\left(\frac{\delta  \mathscr{F}}{\delta C\dopb} - \frac{\delta  \mathscr{F}}{\delta C\dopc}\right) \text{d}V.
\end{equation}

Finally, the weak form of Eq \eqref{PoissonEquation} is
\begin{equation}
    R_\phi = \int_{V}  \epsilon_0\epsilon_\text{r} \phi_{,i}\psi_{\phi,i}\text{d}V - \int_\Gamma \epsilon_0\epsilon_\text{r} \phi_{,i}\psi_{\phi}\hat{n}_i\text{d}\Gamma -  \int_V \psi_\phi \frac{\mathcal{F}}{V_\text{m}}(z\V C\V+z\dop C\dop)\text{d}V.
\end{equation}

Following the Galerkin approach \cite{brenner2008mathematical}, the test functions are discretized as  $\psi_{C\Vc} = N_{C\Vc}^I\psi_{C\Vc}^I$,  $\psi_{C\dopc} = N_{C\dopc}^I\psi_{C\dopc}^I$, $\psi_{\eta_i}=N_{\eta_i}^I\psi_{\eta_i}^I$, $\psi_{C\V}=N_{C\V}^I\psi_{C\V}^I$, $\psi_{C\dop}=N_{C\dop}\psi_{C\dop}^I$, $\psi_{\mu\V}=N_{\mu\V}\psi_{\mu\V}^I$, $\psi_{\mu\dop}=N_{\mu\dop}\psi_{\mu\dop}^I$, $\psi_{C\Vb}=N_{C\Vb}\psi_{C\Vb}^I$, $\psi_{C\dopb}=N_{C\dopb}\psi_{C\dopb}^I$ and $\psi_{\phi} = N_{\phi}\psi_{\phi}^I$, where $I$ denotes the node index and Einstein summation convention is used. $\psi_{C\Vc}^I$,  $\psi_{C\dopc}^I$, $\psi_{\eta_i}^I$, $\psi_{C\V}^I$, $\psi_{C\dop}^I$, $\psi_{\mu\V}^I$, $\psi_{\mu\dop}^I$, $\psi_{C\Vb}^I$, $\psi_{C\dopb}^I$ and $\psi_{\phi}^I$ are the nodal weights, while  $N_{C\Vc}^I$,  $N_{C\dopc}^I$, $N_{\eta_i}^I$, $N_{C\V}^I$, $N_{C\dop}^I$, $N_{\mu\V}^I$, $N_{\mu\dop}^I$, $N_{C\Vb}^I$, $N_{C\dopb}^I$ and $N_{\phi}^I$ are the shape functions for the corresponding variables. Then we can write down the discretized residuals as 
\begin{flalign}
    &R_{C\Vc}^I = \int_V N_{C\Vc}^I\{C\V -  [1-h(\eta)]C\Vb - h(\eta) C\Vc \}\text{d}V &\\
    &R_{C\dopc}^I = \int_V N_{C\dopc}^I\{C\dop -  [1-h(\eta)]C\dopb - h(\eta) C\dopc \}\text{d}V &\\
    &\begin{aligned}
    R_{\eta_i}^I &= \int_V \frac{N_{\eta_i}^I}{\Delta t} \text{d}V + L\int_V N_{\eta_i}^I \omega \frac{\partial f^{\text{loc}}(\eta)}{\partial\eta_i} \text{d}V + L\int_V\kappa \eta_{i,j}N_{\eta_i ,j}^I \text{d}V -L\int_\Gamma\kappa\eta_{i,j}N_{\eta_i}^I \hat{n}_j \text{d}\Gamma\\
   &-L\int_V N_{\eta_i}^I \frac{\partial h(\eta)}{\partial \eta_i}\left[ f_\text{b}^{\text{ech}} - f_{\text{c}}^{\text{ech}} - \frac{\partial f_\text{b}^{\text{ech}}}{\partial C\Vb} (C\Vb - C\Vc)-  \frac{\partial f_\text{b}^{\text{ech}}}{\partial C\dopb} (C\dopb - C\dopc) \right]\text{d}V
    \end{aligned}&\\
    &R_{C\V}^I = \int_V N_{C\V}^I \left(\mu\V - \frac{\delta  \mathscr{F}}{\delta C\Vb}\right)    \text{d}V &\\
    &R_{C\dop}^I = \int_V N_{C\dop}^I \left(\mu\dop - \frac{\delta  \mathscr{F}}{\delta C\dopb}\right)    \text{d}V &\\
    & R_{\mu\V}^I=\int_V  \frac{N_{\mu\V}^I}{\Delta t} \text{d}V + \int_V  M\V\mu_{\text{V}_{\text{O}}^{..},i} N_{\mu\V, i}^I\text{d}V - \int_\Gamma  M\V \mu_{V_{O}^{..},i}N_{\mu\V}^I \hat{n}_{i} \text{d}\Gamma &\\
    &R_{\mu\dop}^I=\int_V \frac{ N_{\mu\dop}^I}{\Delta t} \text{d}V + \int_V  M\dop\mu_{\text{dop},i} N_{\mu\dop, i}^I\text{d}V - \int_\Gamma  M\dop\mu_{\text{dop},i}N_{\mu\dop}^I \hat{n}_{i} \text{d}\Gamma &\\
    &R_{C\Vb}^I = \int_V N_{C\Vb}^I\left(\frac{\delta  \mathscr{F}}{\delta C\Vb} - \frac{\delta  \mathscr{F}}{\delta C\Vc}\right) \text{d}V&\\
     &R_{C\dopb}^I = \int_V N_{C\dopb}^I\left(\frac{\delta  \mathscr{F}}{\delta C\dopb} - \frac{\delta  \mathscr{F}}{\delta C\dopc}\right) \text{d}V&\\
     &R_\phi^I = \int_{V}  \epsilon_0\epsilon_\text{r} \phi_{,i} N_{\phi,i}^I\text{d}V - \int_\Gamma \epsilon_0\epsilon_\text{r} \phi_{,i} N_{\phi}^I\hat{n}_i\text{d}\Gamma -  \int_V N_\phi^I \frac{\mathcal{F}}{V_\text{m}}(z\V C\V+z\dop C\dop)\text{d}V &
\end{flalign}
The linearization of the residuals can be written in the element-level linear formulation, in which the terms in the tangent matrix are calculated by $K_{\xi,\zeta}^{IJ} = \partial R_{\xi}^I/\partial \zeta^J$, where ${\xi, \zeta}$ indicates individual DOF. The non-zero terms in the tangent  matrix are presented in supplementary. The numerical implementation of the proposed phase-field model is carried out in Multiphysics Object-Oriented Simulation Environment (MOOSE) framework \cite{schwen2023phasefield, giudicelli2024moose}. {Spatial discretization was performed with linear elements for one-dimensional (1D) simulations (Section 5.1 and 5.2) and quadrilatera elements for two-dimensional (2D) simulations (Section 5.3), with linear Lagrange shape functions in each case \cite{aagesen2024electrochemical}. The same basis functions were used for both trial (shape) and test functions, which is standard in the Galerkin finite element method \cite{brenner2008mathematical}. Different mesh resolutions were adopted based on the specific requirements of each case. To ensure numerical accuracy, the relative convergence tolerance is set to $1 \times 10^{-8}$. All phase-field simulations were conducted on CPU nodes equipped with Intel Xeon Platinum 8470Q processors and were accelerated using the Message Passing Interface (MPI) on the Lichtenberg high-performance computing cluster at Technische Universit\"at Darmstadt.}


\section{Phase-field simulation results of Fe-doped SrTiO$_3$}
\label{FEM_phasefield_simulations}

 We present first in \Cref{equilibrium} one-dimensional (1D) phase-field simulation results to reproduce the formation of a SCL at equilibrium state {under the influence of different defect chemistry parameters}.  Subsequently, we investigate the influence of solute drag effects induced by the segregation of point defects in the core on the formation of an asymmetric SCL under quasi-equilibrium state in \Cref{quasi-equilibrium}. Finally, we conduct two-dimensional (2D) phase-field simulations to study the microstructure evolution process in STO polycrystalline material during sintering and analyze the resulting grain size distribution in \Cref{2Dmicrostrucutre}. 

Note that in the paper we ignore the GB orientation dependency or other anisotropy of the segregation energies $\Delta g\V$ and $\Delta g\dop$, for we intend to reveal the most fundamental mechanisms of SCL formation and the grain growth patterns. Without loss of generality, the proposed phase-field model can be extended accordingly to include such anisotropy aspects.

\subsection{GB space charge layers at equilibrium} \label{equilibrium}

{To examine the influence of various defect parameters on SCL formation, 1D phase-field simulations are conducted for a bicrystal configuration at temperatures of $T = 600$ K and $T = 1623$ K. At 600 K, the distribution of acceptor dopant is frozen, while that of the oxygen vacancy achieves chemical equilibrium (MS model). At 1623 K, the diffusion of acceptor dopant becomes active, enabling both the acceptor dopant and oxygen vacancy to achieve electrochemical equilibrium, referred to as the GC model. Furthermore, we chose the physical GB width (0.78 nm \cite{de2009formation}) in the phase-field simulations. This choice can decrease the abnormal interface effects raised by an artificially enlarged interface thickness \cite{wang2020modeling, wang2021quantitative}. The defect chemistry parameters used for the phase-field simulations of SCLs are summarized in \Cref{Equilibrium_SCL_parameters}.}

{We summarize the numerical settings employed in the FEM implementation. The simulation domain is one-dimensional and symmetric, spanning from $x = -390$ nm to $x = 390$ nm, with the GB located at $x = 0$ nm. The domain is discretized into 10,000 elements with a uniform element size of 0.078 nm.
For the electrostatic potential, a Neumann boundary condition is applied at the right boundary, while the left boundary is grounded. Neumann boundary conditions are also imposed on the concentration field at both boundaries.  Each simulation is executed on a single CPU node with 16 cores and 8 gigabyte of memory, requiring approximately 5 hours of computation time.}

\begin{table}
\caption{{The defect chemistry parameters for the 1D phase-field simulations in MS and GC models to reproduce equilibrium SCLs. The definition of these parameters are described in \Cref{shape interface parameters}.}}
\label{Equilibrium_SCL_parameters}
    \centering
    {
    \begin{tabular}{lcc}
    \hline 
        Parameters  & Value in MS model \cite{de2009formation} & Value in GC model \cite{de2009formation} \\
        \hline 
         $T$/K  & 600 & 1623 \\
         $a$/nm & 0.39& 0.39 \\
         $\omega_c$/nm & 0.78& 0.78 \\
         $\epsilon_r$   &160 & 56 \\
         $c\dopb (\infty)$/ $\text{m}^{-3}$ & $7\times 10^{25}$ & $7\times 10^{25}$\\
         $c\Vb (\infty)$/ $\text{m}^{-3}$  & $3.5\times 10^{25}$ & $3.5\times 10^{25}$\\
         $l_\text{D}$/nm & 1.81 & 1.76\\ 
         $\Delta g\V$/eV  & -1.5 $\sim$ 0  & -1.5\\
         $\Delta g\dop$/eV & - &  -1.0 $\sim$ 0.5\\
         $N\Vb$/ $\text{m}^{-3}$ & $5.1\times 10^{28}$  &  $5.1\times 10^{28}$ \\
         $N\dopb$/ $\text{m}^{-3}$  &  -  &  $1.68\times 10^{28}$ \\
          $N\Vc$/ $\text{m}^{-3}$ & $5\times 10^{26} $ $\sim$ $1\times 10^{27} $ &  $5\times 10^{26} $ $\sim$ $1\times 10^{27} $ \\
         $N\dopc$/ $\text{m}^{-3}$ & -  &  $1.68\times 10^{26}$ $\sim$ $1.68\times 10^{27}$ \cite{de2009formation}\\
         \hline 
    \end{tabular}
    }
\end{table}

\subsubsection{Benchmark case 1: MS Model}
  By prescribing a uniform dopant concentration, the proposed phase-field model is employed to simulate the MS case. {Phase-field simulations of SCLs with varying available number densities in the MS model are systematically benchmarked against the sharp interface (SI) calculations of GB$|$SCL configurations, as described in \Cref{defect chemistry}. The results of this comparison are presented in \Cref{fig3_new}. The phase-field results are represented by solid lines, while the SI calculations are denoted by cross symbols. The orange, blue, and purple colors correspond to different available site densities: orange for $N\Vc=1\times 10^{27} \text{m}^{-3}$, blue for $N\Vc=7.5\times 10^{26} \text{m}^{-3}$, and purple for $N\Vc=5\times 10^{26} \text{m}^{-3}$.}

{\Cref{fig3_new}(a) and (b) illustrate the GB potentials ($\Phi_0$) and oxygen vacancy concentrations at the GB core ($c\Vc$) for various segregation energies and available oxygen vacancy number densities. It should be clarified that $\Phi_0$ and $c\Vc$ are directly obtained at the interface position, defined by $\eta = 0.5$ in the phase-field method. An increase in $|\Delta g\V|$ ($\Delta g\V < 0$) promotes greater oxygen vacancy segregation at the GB core, thereby enhancing the GB potential. However, the GB potential does not increase indefinitely. At sufficiently large $|\Delta g\V|$, when all available sites for oxygen vacancies are fully occupied, the GB potential reaches a plateau, as does the oxygen vacancy concentration in the GB core ($c\Vc \approx N\Vc$, see \Cref{fig3_new}(b)). The available number density of oxygen vacancy sites ($N\Vc$) plays a critical role in determining the plateau value.
For instance, when $N\Vc$ increases from $5 \times 10^{26} \text{ m}^{-3}$ to $7.5 \times 10^{26} \text{ m}^{-3}$, the plateau GB potential rises from 0.12 V to 0.24 V in the phase-field results and from 0.13 V to 0.27 V in the sharp interface (SI) calculations. However, when $N\Vc$ is as high as $1 \times 10^{27} \text{ m}^{-3}$, this plateau behavior becomes less pronounced, as no clear convergence of the GB potential is observed in both the SI calculations and phase-field results. Although phase-field results exhibit the same trend as predicted by SI calculations, discrepancies in GB potentials are still observed in the large $|\Delta g\V|$ region.}

{To further investigate the GB core, we plot the area density of charge in the GB core, $Q_\text{c} = \int_{-w_c/2}^{w_c/2} e(z\dop c\dopc + z\V c\Vc)\text{d}x$, obtained from phase-field results and SI calculations, as a function of $\Delta g\V$ for different $N\Vc$ in \Cref{fig3_new}(c). Here, $Q_\text{c}$ represents the boundary condition in the SI calculations, as defined in \Cref{BC_left}, which plays a critical role in determining the GB potential and defect concentration in GB core.
When $|\Delta g\V|$ is low, the phase-field results align well with the SI calculations. However, discrepancies emerge in the high $|\Delta g\V|$ region, where phase-field simulations reproduce lower area charge densities. These discrepancies are likely due to the diffuse interface in phase-field simulations. In the SI model, the defect concentrations between the GB core and the bulk phase are discontinuous, with defect concentrations assumed to be constant across the GB core. Conversely, the smooth interface characteristics inherent to phase-field simulations result in a less positively charged GB core compared to the SI calculations when the same GB width, $w_c = 0.78$ nm, is employed.}

{In addition to the GB core, we also compare the distributions of electrostatic potentials and oxygen vacancy concentrations in the bulk phase for different $\Delta g\V$ values, as shown in \Cref{fig3_new}(d) to (f). Due to the symmetric bicrystal configuration, only half of the interface region is included in the analysis.
In \Cref{fig3_new} (d1), (e1), and (f1), depletion zones of oxygen vacancy are observed near the GB core. Larger $N\Vc$ values result in more pronounced depletion of oxygen vacancy. Overall, the phase-field results show good agreement with the SI calculations. n \Cref{fig3_new}(d1), (e1), and (f1), depletion zones of oxygen vacancy are observed near the GB core. Larger $N\Vc$ values lead to more pronounced depletion of oxygen vacancy. Overall, the phase-field results align well with the SI calculations. However, discrepancies are observed near the GB core, where oxygen vacancy concentrations decrease in the SI calculations but increase in the phase-field results (see \Cref{fig3_new}(e1) and (e2) in the range of [0, 2 nm]). These differences stem from the contrasting assumptions used to connect the GB core and bulk phase in the phase-field model and the discontinuous SI model, as discussed earlier.
In \Cref{fig3_new} (d2), (e2), and (f2), larger $N\Vc$ valules lead to higher electrostatic potential and longer electrostatic decay lengths. The phase-field simulations is capable of capturing these trends and nicely reproducing the SI predictions.}

\begin{figure}
    \centering
    \includegraphics[width=0.9\linewidth]{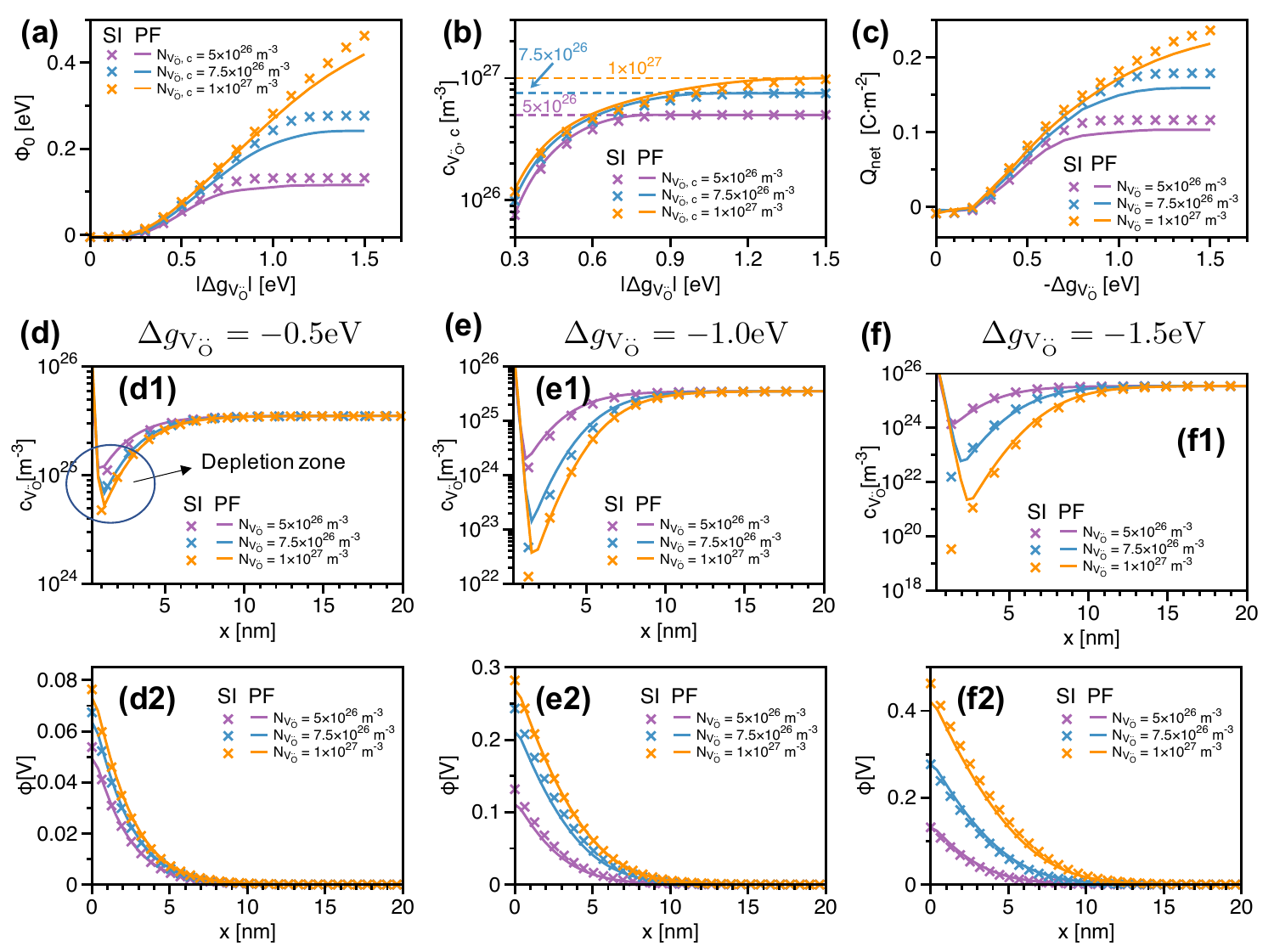}
    \caption{(a) Comparison of phase-field (PF) simulation outcomes with those of the abrupt GB$|$SCL sharp interface (SI) predictions with different available number density of oxygen vacancy, such as $N\Vb =5\times 10^{26} \text{m}^{-3}$, $7.5\times 10^{26} \text{m}^{-3}$ and $1\times 10^{27} \text{m}^{-3}$. Figure (a), (b) and (c) illustrates the GB potential, GB concentration of oxygen vacancy and area density of charge as a function of $\Delta g\V$. Figure (d), (e) and (f) depicts the spatial distribution of oxygen vacancy concentration and electrostatic potential  in bulk phase.}
    \label{fig3_new}
\end{figure}

\subsubsection{Benchmark case 2: GC Model}
By fixing the electrochemical potentials of the {acceptor} dopant and oxygen vacancies as constants, the proposed phase-field model reduces to the GC model, enabling the reproduction of equilibrium SCL formation in the GC case. {In this framework, the effects of the segregation energy of the acceptor dopant ($\Delta g\dop$) and the available site densities of oxygen vacancies and acceptor dopants in the GB core ($N\dopc$ and $N\Vc$) on SCL formation are investigated and compared with SI calculations, as illustrated in \Cref{fig5_new}. In both phase-field and SI results, the segregation energy of oxygen vacancy $\Delta g\V$ is set to 1.5 eV. For $N\Vc=1\times 10^{27} \text{m}^{-3}$, phase-field results and SI predictions are shown using solid lines and cross symbols, respectively. For $N\Vc=5\times 10^{26} \text{m}^{-3}$, dashed lines and triangle symbols are used. Additionally, different colors are employed to represent varying $N\dopc$: red for $N\dopc = 1.68\times 10^{26} \text{m}^{-3}$, blue for $N\dopc = 8.4\times 10^{26} \text{m}^{-3}$, and orange for $N\dopc = 1.68\times 10^{27} \text{m}^{-3}$.}

{In \Cref{fig5_new}(a), (b), and (c), the GB potential ($\Phi_0$), GB concentrations of oxygen vacancy ($c\Vc$), and acceptor dopant ($c\dopc$) are presented as functions of $\Delta g\dop$. In \Cref{fig5_new}(a), the GB potential curves exhibit two distinct plateaus. For positive $\Delta g\dop$, the GB potential reaches an upper plateau due to minimal dopant segregation in the GB core. In this case, the dopant concentration in the GB core can be 2 to 3 orders of magnitude lower than in the bulk phase (see \Cref{fig5_new}(b) at positive $\Delta g\dop$ region), making it insufficient to compensate for the charge of the GB core induced by oxygen vacancy. As $\Delta g\dop$ becomes negative, the positively charged GB core is gradually compensated by the segregating dopant (see \Cref{fig5_new}(b) at negative $\Delta g\dop$ region), leading to a reduction in GB potential. When all available sites for acceptor dopant are fully occupied, the GB potential reaches its lower plateau. 
Increasing $N\Vc$ raises both the upper and lower plateaus simultaneously, with little impact on the shape of the GB potential curve. The difference between the upper and lower plateaus is primarily determined by $N\dopc$, with larger $N\dopc$ resulting in a greater plateau difference. For example, when $N\Vc = 5\times 10^{26} \text{m}^{-3}$, the plateau difference is 0.17 V for $N\dopc = 1.68\times 10^{26} \text{m}^{-3}$ and increases to 1.5 V for $N\dopc = 1.68\times 10^{27} \text{m}^{-3}$. In addition, very large $N\dopc$ might lead to negative GB potential as presented by the orange dashed line and triangle symbols when $\Delta g\dop$ is smaller than -0.8 eV in \Cref{fig5_new}(a). }

{Adjusting $N\Vc$ and $N\dopc$ has a pronounced effect on the GB concentrations of oxygen vacancy and acceptor dopant. When the available sites are not fully occupied, increasing $N\Vc$ and $N\dopc$ results in greater dopant accumulation within the GB core for a given $\Delta g\dop$, as shown in \Cref{fig3_new}(b). The segregation of dopant also influences the distribution of oxygen vacancy. In \Cref{fig3_new}(c), a more negative $\Delta g\dop$ leads to a higher GB concentration of oxygen vacancy. Similar to the GB potential curves, the oxygen vacancy concentration curves exhibit two plateaus. An increase in $N\dopc$ amplifies the difference between these plateaus.}

{We also present the distributions of electrostatic potential, oxygen vacancy concentration, and acceptor dopant concentration in the bulk phase for different $N\dopc$ and $\Delta g\dop$ in \Cref{fig5_new}(d), (e), and (f). Here, $N\Vc$ is fixed at $5\times 10^{26} \text{m}^{-3}$. First, we observe a depletion zone for oxygen vacancy and an accumulation zone for acceptor dopant in the SCLs when the GB potential is positive, while the opposite behavior is observed for negative GB potential (see the orange dashed lines and triangle symbols in \Cref{fig5_new}(f1), (f2), and (f3)).
For $\Delta g\dop = 0$ eV, increasing $N\dopc$ has minimal impact on the distributions of oxygen vacancy concentration, acceptor dopant concentration, and electrostatic potential in the bulk phase. As a result, the curves for $N\dopc = 1.68\times 10^{26} \text{m}^{-3}$, $8.4\times 10^{26} \text{m}^{-3}$, and $1.68\times 10^{27} \text{m}^{-3}$ are nearly indistinguishable, as shown in \Cref{fig5_new}(d1), (d2), and (d3).
For $\Delta g\dop = -0.5$ eV and $-1.0$ eV, increasing $N\dopc$ raises the oxygen vacancy concentration in the depletion zone, reduces the acceptor dopant concentration in the accumulation zone, and decreases the electrostatic potential in SCLs. However, we still observe discrepancies between phase-field results and SI predictions near the GB core, which has been clarified in the MS case.}

{We compare the equilibrium SCL formation obtained from phase-field simulations with the results of Vikrant's simulations for Fe-doped SrTiO$_3$ (see Figure 1 in Ref. \cite{vikrant2020electrochemical}). In Vikrant's study, {a higher dopant concentration is used, i.e. 2\% (in present work, dopant concentration is 0.42\%), which is considered as non-dilute case and different grain boundary misorientation is considered. Their simulation temperature is set at 1623 K, and the acceptor dopant is considered mobile. They employed a negative segregation energy for oxygen vacancy ($-0.8$ eV) and a zero segregation energy for dopant. }

The major differences between the two studies are as follows: (i) In Vikrant's results, the dopant concentration in the GB core is consistently higher than in the bulk phase for all GB misorientations.
(ii) Their simulations also reveal a depletion zone for acceptor dopant near the GB core within the SCL.
 In the present work, a zero segregation energy for dopant results in significantly lower dopant concentrations in the GB core. As shown in \Cref{fig5_new}(b), the dopant concentration in the GB core, $c\dopc$, is consistently smaller than the bulk concentration, $c\dopb(\infty) = 7\times 10^{25}$ $\text{m}^{-3}$, for all cases where $\Delta g\dop = 0$ eV. Furthermore, we observe an accumulation zone for acceptor dopant in the SCL region under positive electrostatic potentials, as depicted in \Cref{fig5_new}(d2), (e2) and (f2). 

 {We note that defect–defect interactions were included in Vikrant’s work and can influence space charge layer formation, as discussed in Refs. \cite{lund2021thermodynamically, mebane2015generalised}. However, their model assumes that oxygen vacancies and dopants share the same sublattice, which affects the configurational entropy and consequently the electrochemical potential (see Section 3.3), particularly in the non-dilute regime where these quantities are sensitive to the underlying sublattice occupancy assumptions.}
 

{Thus far, we have successfully benchmarked the capabilities of the proposed phase-field model against the abrupt GB$|$SCL model in both the MS and GC cases. By comprehensively accounting for the segregation energies of various defects and partitioning the available site densities among different defects, diverse SCL formation behaviors have been accurately reproduced.
As a result, the phase-field results show excellent agreement with SI predictions, not only within the GB core but also in the bulk phase. However, due to the assumptions of discontinuous GB and SCL in the SI model, two key discrepancies are observed. First, the GB potential in the GB core obtained from the phase-field model is smaller than that predicted by the SI model, particularly when the segregation energy is sufficiently large. Second, differences in defect concentrations are confined to the immediate vicinity of the GB core.}

\begin{figure}
    \centering
   \includegraphics[width=0.9\linewidth]{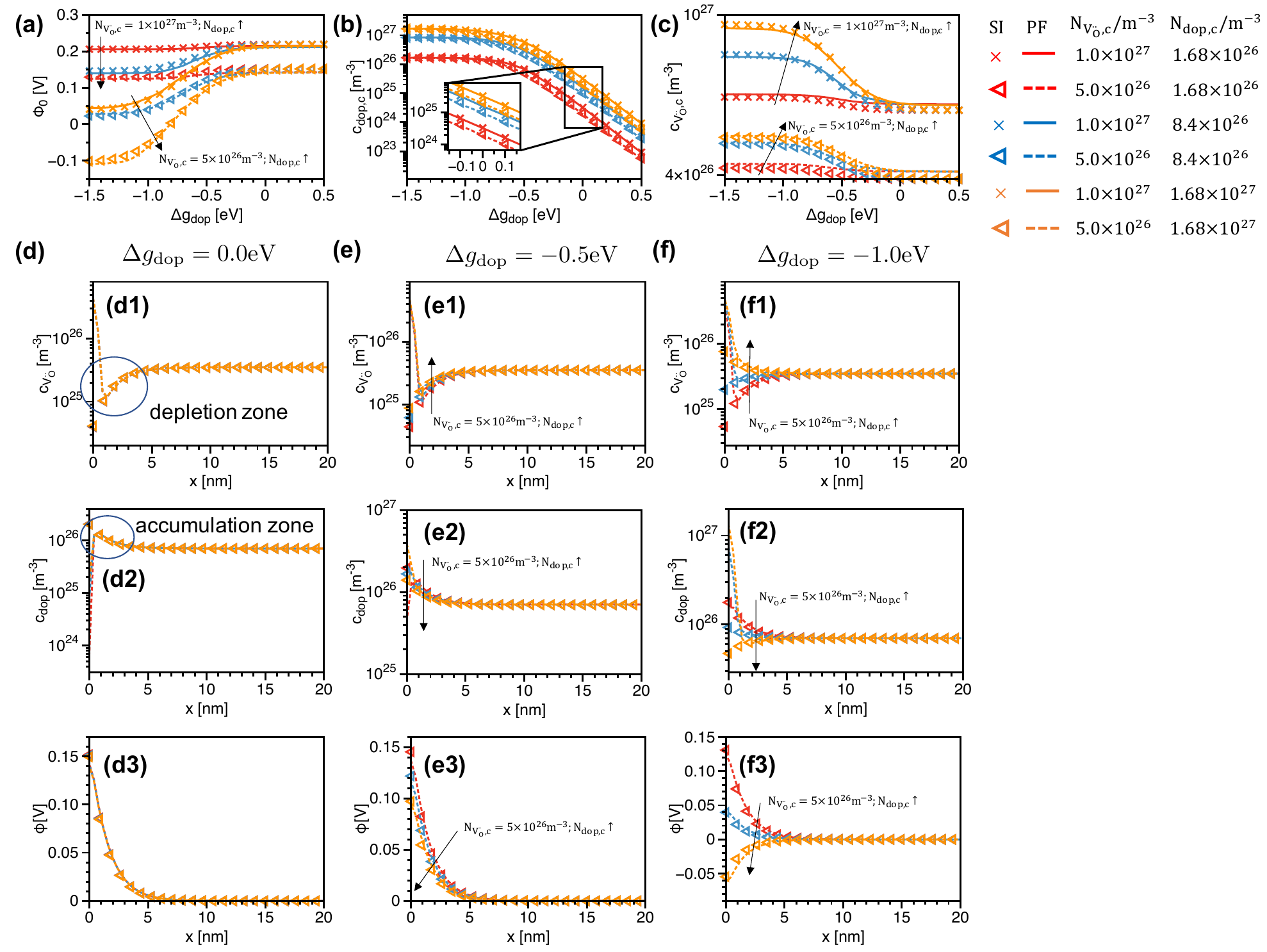}
    \caption{Comparison of phase-field (PF) simulation results with the abrupt GB$|$SCL sharp interface (SI) results via FEM method for the GC case. Figures (a), (b) and (c) showcase the variations in grain boundary potential, dopant concentration, and oxygen vacancy concentration in the core as functions of $\Delta g_{\text{dop}}$. Figures (d), (e), and (f) detail the spatial distributions of dopant and oxygen vacancy concentrations in the bulk phase and the electrostatic potential.  {$\Delta g\V$ was kept at -1.5 eV for all calculations}.}
    \label{fig5_new}
\end{figure}

\subsection{GB space charge layers at quasi-equilibrium} \label{quasi-equilibrium}

In this section, we systematically investigate the solute drag effect induced by an asymmetric SCL at a quasi-statically moving GB in a STO bicrystal case. {This analysis is conducted using the defect-chemistry-consistent phase-field model, which has been validated against sharp interface predictions in previous section. While the impacts of GB misorientation have been extensively studied in Ref. \cite{vikrant2020electrochemical}, our focus here is on the influence of defect chemistry parameters, such as the segregation energies of oxygen vacancy and dopant, on SCL formation and solute drag effects at the quasi-equilibrium state in a bicrystal case. } 

{The phase-field simulation parameters to reproduce the SCLs at quasi-equilibrium state are listed in \Cref{quasi_Equilibrium_SCL_parameters}. For FEM implementation, the simulation domain is a 1D space extending from $x = -780$ nm to $x = 780$ nm, with the GB core initially positioned at $x = -390$ nm ($t = 0$). The domain size is chosen to be sufficiently large to ensure the movement of SCL is not hindered by the boundary of simulation box. The domain is discretized into 20,000 elements, each with a uniform size of 0.078 nm.
For the electrostatic potential, a Neumann boundary condition is applied at the right boundary, while the left boundary is grounded. Similarly, Neumann boundary conditions are imposed on the concentration field at both boundaries. Each simulation is performed on a single CPU node with 16 cores and 8 GB of memory, requiring approximately 12 hours to reach the quasi-equilibrium state.} 

\begin{table}
\caption{{The defect chemistry parameters for the 1D phase-field simulations  to reproduce quasi-equilibrium SCLs. }}
\label{quasi_Equilibrium_SCL_parameters}
    \centering
    {
    \begin{tabular}{lc}
    \hline 
        Parameters  & Value   \\
        \hline 
         $T$/K  & 1623 \\
         $D\V$/$\text{m}^{-2}\text{s}^{-1}$ & $1.35\times 10^{-9}$\\
         $D\dop$/$\text{m}^{-2}\text{s}^{-1}$ & $1.3\times 10^{-11}$\\
         $a$/nm &  0.39 \\
         $\omega_c$/nm &  0.78 \\
         $\epsilon_r$   & 56 \\
         $c\dopb (\infty)$/ $\text{m}^{-3}$ & $2\times 10^{25}$ $\sim$ $2\times 10^{26}$ \\
         $c\Vb (\infty)$/ $\text{m}^{-3}$  & $0.5\times c\dopb (\infty)$ \\
         $\Delta g\V$/eV  & -0.5 $\sim$ -1.75\\
         $\Delta g\dop$/eV &   -0.2 $\sim$ -0.7\\
         $N\Vb$/ $\text{m}^{-3}$ & $5.1\times 10^{28}$   \\
         $N\dopb$/ $\text{m}^{-3}$   &  $1.68\times 10^{28}$ \\
          $N\Vc$/ $\text{m}^{-3}$ & $1\times 10^{27} $ \\
         $N\dopc$/ $\text{m}^{-3}$  &  $1.68\times 10^{27}$ \\
         \hline 
    \end{tabular}
    }
\end{table}

{To investigate the formation of asymmetric SCLs induced by solute drag effects under quasi-equilibrium conditions, a constant core velocity (${v}_{\text{c}}$) is prescribed as a parameter in the simulations. The core velocity ranges from 0.01 $\text{nm}\,\text{s}^{-1}$ to 100 $\text{nm}\,\text{s}^{-1}$ to study its influence on the SCL. The simulation results, shown in \Cref{fig6}, illustrate the distributions of acceptor dopant concentration, oxygen vacancy concentration, charge density, and electrostatic potential as functions of $x$. The charge density is calculated as $\rho(x) = e(z\V c\V + z\dop c\dop)$. The GB core moves from left to right along the $x$-axis.
The acceptor dopant plays a crucial role in solute drag effects due to its significantly lower diffusivity compared to oxygen vacancy. Adjusting the segregation energy of the dopant significantly alters its concentration distribution. To examine the impact of segregation behavior, two segregation energies for the acceptor dopant, $0$ eV and $-0.5$ eV, are selected to generate different SCLs. The segregation energy of oxygen vacancy is fixed at $-1.5$ eV.
In the following discussion, the region to the left of the GB core is referred to as the growing grain, while the region to the right is referred to as the shrinking grain.}

{The SCLs at the equilibrium state (i.e., $v_\text{c}$ = 0 $\text{nm}\,\text{s}^{-1}$) are shown in \Cref{fig6}(a) as a reference. The two different segregation energies for the acceptor dopant lead to distinct concentration distributions, as illustrated in \Cref{fig6}(a1). For $\Delta g\dop = 0$ eV, the acceptor dopant concentration exhibits a depression at the GB core, while $\Delta g\dop = -0.5$ eV results in segregation. Moreover, the charge density in the GB core is higher for $\Delta g\dop = 0$ eV than for $\Delta g\dop = -0.5$ eV. Consequently, the GB potential for $\Delta g\dop = -0.5$ eV is smaller than that for $\Delta g\dop = 0$ eV.}

{Compared with symmetric SCLs, a moving GB core introduces notable differences. First, the concentration profiles of the dopant are highly sensitive to GB velocity. The symmetric pattern begins to deteriorate even at very low velocities, such as 0.01 $\text{nm}\,\text{s}^{-1}$, as shown in \Cref{fig6}(b1). As a result, the dopant accumulation zone increases in the growing grain but decreases in the shrinking grain. At higher core velocities, dopant cations fail to keep up with the movement of the GB core, leading to a significant decrease in dopant concentration within the core, as illustrated in \Cref{fig6}(c1), (d1), and (e1). When the core velocity reaches $v_\text{c}$ = 100 $\text{nm}\,\text{s}^{-1}$, the dopant concentration profile becomes nearly flat, as depicted in \Cref{fig6}(e1).
Second, significant asymmetry in the oxygen vacancy concentration arises only at very high GB velocities, due to its high diffusivity. As depicted in \Cref{fig6}(b2), (c2), and (d2), the oxygen vacancy concentration profiles exhibit only slight asymmetry at lower velocities. However, when $v_\text{c}$ = 100 $\text{nm}\,\text{s}^{-1}$, the concentration of oxygen vacancy in the depletion zone of the growing grain increases notably, as shown in \Cref{fig6} (e2). Simultaneously, the depletion zone in the growing grain becomes substantially elongated, whereas it contracts in the shrinking grain. Third, as the GB velocity increases, not only does the GB potential rise, but a difference in electrostatic potential between the growing and shrinking grains also emerges, as shown in \Cref{fig6}(b4)–(e4). Additionally, the decay length of the electrostatic potential in the growing grain becomes significantly longer when GB velocity increases, resulting in an elongated SCL length in the growing grain.}

{The segregation energy of the acceptor dopant plays a crucial role in SCL formation when the GB core is moving. Different values of $\Delta g\dop$, namely $-0.5$ eV and $0$ eV, lead to distinct dopant concentration distributions. In \Cref{fig6}(b1), when the GB velocity $v_\text{c}$ is $0.01$ $\text{nm}\,\text{s}^{-1}$, the accumulation zones in the SCL for both segregation energies exhibit the same trend. However, as $v_\text{c}$ increases to $0.5$ $\text{nm}\,\text{s}^{-1}$ and 2 $\text{nm}\,\text{s}^{-1}$, notable differences emerge as shown in \Cref{fig6}(c1) and (d1). For $\Delta g\dop = -0.5$ eV, the accumulation zone in the shrinking grain transforms into a depletion zone, where the dopant concentration becomes lower than in the bulk. Meanwhile, the accumulation zone in the growing grain persists. In contrast, for $\Delta g\dop = 0$ eV, the opposite behavior is observed. The accumulation zone remains in the shrinking grain, while in the growing grain, it diminishes into a depletion zone. When $v_\text{c}$ = 100 $\text{nm}\,\text{s}^{-1}$, the dopant concentration becomes flat for both cases.
The influence of the segregation energy of the acceptor dopant on the electrostatic potential is significant only at low GB velocities. A more negative $\Delta g\dop$ leads to a smaller GB potential. At high GB velocities, where the dopant concentration profiles flatten, their effect on the electrostatic potential distribution diminishes, as shown in \Cref{fig6}(c4), (d4), and (e4).}

\begin{figure}
    \centering
    \includegraphics[width=1\linewidth]{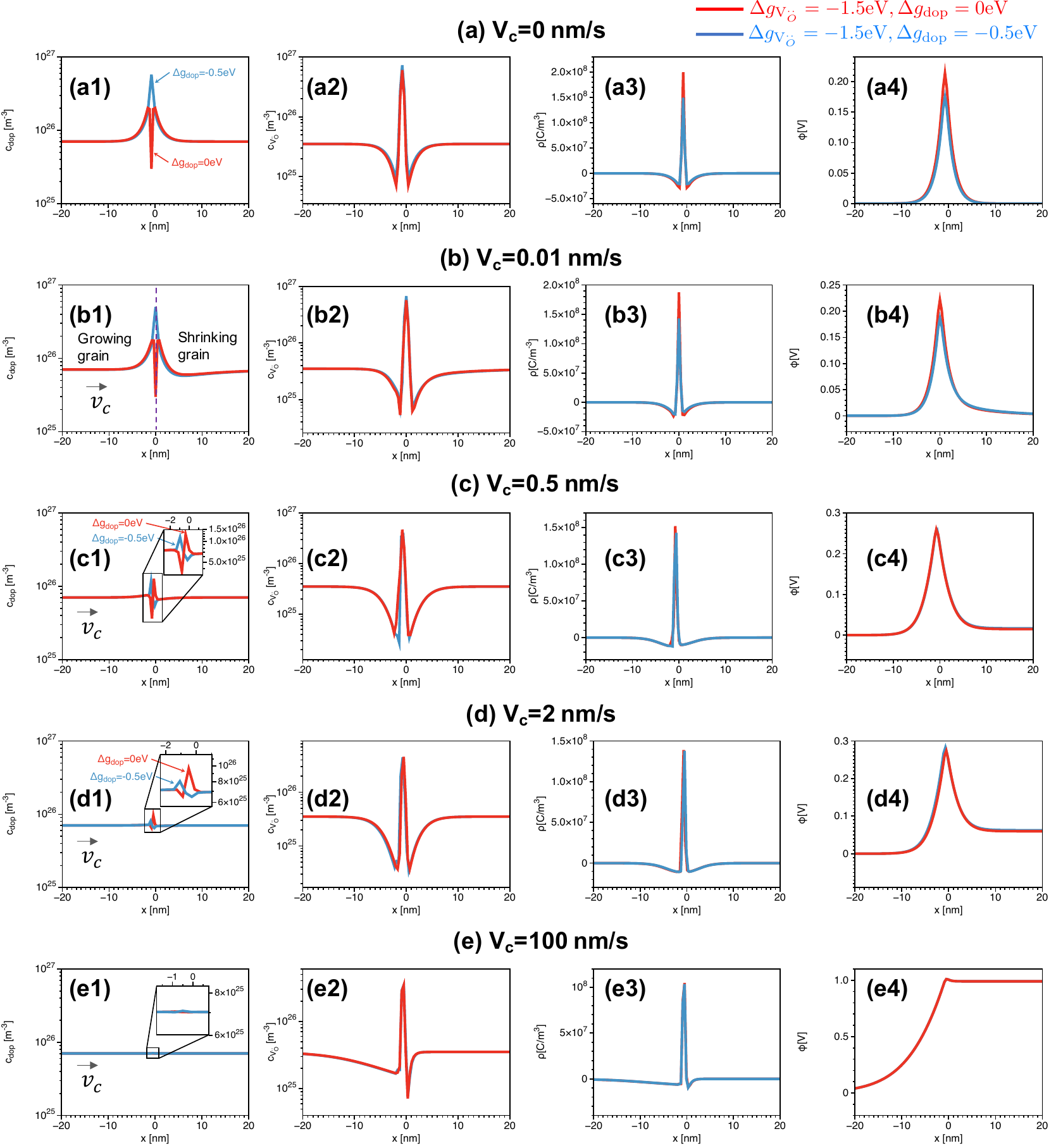}
    \caption{Asymmetric SCL formation in a STO bicrystal system with $c\dopb(\infty) = 7.0\times 10^{25} $ $ \text{m}^{-3}$,  $\Delta g\V=-1.5$ eV and two $\Delta g\dop=$ 0 and -0.5 eV at quasi-equilibrium state under different constant core velocity (a) $v_\text{c} = 0$  $\text{nm}\,\text{s}^{-1}$, (b) $v_\text{c} = 0.01$ $\text{nm}\,\text{s}^{-1}$, (c) $v_\text{c} = 0.5$ $\text{nm}\,\text{s}^{-1}$, (d) $v_\text{c} = 2$ $\text{nm}\,\text{s}^{-1}$, (e) $v_\text{c} = 100$ $\text{nm}\,\text{s}^{-1}$. Each row has four sub-figures, depicting the distribution of dopant concentration, oxygen vacancy concentration, total charge density and electrostatic potential, respectively. In this figure, blue line indicates $\Delta g\dop$ = -0.5 eV, while red line is $\Delta g\dop$ = 0 eV. }
    \label{fig6}
\end{figure}

{The influence of defect chemistry parameters, especially segregation energy of dopant, on the formation behavior of different SCLs has been demonstrated.} In order to further investigate solute drag effects and identify the critical core velocity at which a velocity jump occurs during sintering process, we present the core velocity as a function of the total driving force calculated via the following equation taken from Ref. \cite{vikrant2020electrochemical}
\begin{equation} \label{drag_force1}
\begin{split}
F_\text{T} &= \int_{-\infty}^{\infty} \frac{v_c}{L}\left(\frac{\partial \eta}{\partial x}\right)^2 \text{d}x + \int_{-\infty}^{\infty} {\mu\V}\frac{\partial C\V}{\partial x}\text{d}x +  \int_{-\infty}^{\infty} {\mu\dop}\frac{\partial C\dop}{\partial x}\text{d}x \\
& = \int_{-\infty}^{\infty} \frac{v_c}{L}\left(\frac{\partial \eta}{\partial x}\right)^2 \text{d}x - \int_{-\infty}^{\infty} \left[C\V - C\Vb(\infty)\right]\frac{\partial \mu\V}{\partial x} \text{d}x - \int_{-\infty}^{\infty} \left[C\dop - C\dopb(\infty)\right]\frac{\partial \mu\dop}{\partial x} \text{d}x.
\end{split}
\end{equation}
The second and third terms in \Cref{drag_force1} denote as the electrochemical drag force, which accounts for the electrostatic and chemical contributions when the expressions of electrochemical potential is employed. The integration in \Cref{drag_force1} are solved numerically.
 
The segregation energy difference, $\Delta g\dop$ and $\Delta g\V$, allow direct tuning of the concentrations of oxygen vacancy and dopant in the core, thereby impacting the critical total driving force required for a specific core velocity.
In \Cref{fig7}(a), we hold $\Delta g\dop$ and $c\dopb(\infty)$ constant at -0.5 eV and $7\times10^{25}$ $  \text{m}^{-3}$ and progressively decrease $\Delta g\V$ from -0.5 eV to -1.75 eV. {During the grain shrinking process, the driving force resulting from the GB curvature becomes increasingly significant. Consequently, the velocity of the GB core evolves from a low value to a high value.}
At low core velocities, a more negative $\Delta g\V$ induces higher concentration of oxygen vacancy in the core consequently demanding a larger driving force to maintain the same velocity. 
As the core velocity increases to 1 $\text{nm}\,\text{s}^{-1}$, the dopant {cations} are unable to keep pace with the core movement, resulting in a nearly flat dopant concentration within the core. Hence, this leads to the same linear relationship between the total driving force and the core velocity for different $\Delta g\V$ at high velocity.
On the other hand, for fixed $\Delta g\V$, such as $\Delta g\V$=-1.75 eV, we clearly observe the velocity jumps from 0.065 $\text{nm}\,\text{s}^{-1}$ to approximate 1.5 $\text{nm}\,\text{s}^{-1}$ at $1.45\times 10^{7}$ $ \text{N}\text{m}^{-2}$. As we increase $\Delta g\V$ to -1.0 eV, we hardly observe the velocity jump. When $\Delta g\V$ = -0.5 eV, both the concentrations of dopant and oxygen vacancy in the core decrease significantly. The solute drag effect therefore becomes negligible. {Therefore, decreasing the segregation energy of oxygen vacancy could be an effective way to eliminate the solute drag effects.}
\Cref{fig7}(b) shows the influence of $\Delta g\dop$ on solute drag effects when $\Delta g\V$ and $c\dopb$ are held constant at -1.5 eV and $7\times10^{25}$ $\text{m}^{-3}$. In contrast to \Cref{fig7}(a), we can evidently observe velocity jumps for all $\Delta g\dop$. {When the segregation energy of dopant decrease to -0.2 eV, we can still observe the solute drag effects.}
In \Cref{fig7}(c), we explore the impact of the dopant concentration in the bulk far from the core, denoted as $c\dopb(\infty)$. {We fix $\Delta g\V$ and $\Delta g\dop$ at -1.5 eV and -0.5 eV.}  {We clearly find the solute drag effect is strongly dependent on dopant concentration as evidenced in Ref. \cite{zahler2023grain}. When $c\dopb = 2\times 10 ^{26} \text{m}^{-3}$ (corresponding to 0.12\% site fraction), solute drag effect still remains. }  As $c\dopb (\infty)$ increases, a larger driving force is necessary to counterbalance the solute drag effect to achieve the same core velocity.
 Additionally, a lower critical velocity is observed when the velocity jump occurs at smaller values of $c\dopb (\infty)$. Specifically, the critical velocity increases from 0.065 $\text{nm}\,\text{s}^{-1}$ to 0.095 $\text{nm}\,\text{s}^{-1}$, as $c\dopb (\infty)$ is raised from $4\times10^{25}$ $\text{m}^{-3}$ to $2\times10^{26}$ $\text{m}^{-3}$.

{In the sintering process, the driving forces are initially large and decrease as the grain size increases. In other words, the GB core velocity starts on the upper right side of the S-shaped curves in \Cref{fig7} at the initial stage of the grain growth. As the grain size increases, the GB core velocity decreases to the critical velocity. At this point, a transition from high velocity to low velocity occurs at the critical value. 
In \Cref{fig7}(a), we observe GB core velocity jumps when $\Delta g\V=$ -1.5 eV and -1.75 eV. The critical velocities for different $\Delta g\V$ are approximately 0.35 $\text{nm}\,\text{s}^{-1}$.  Additionally, the GB core velocity jumps to a lower value when $\Delta g\V$ is more negative. In \Cref{fig7}(b), the critical GB core velocity increases from 0.15 $\text{nm}\,\text{s}^{-1}$ to 0.65 $\text{nm}\,\text{s}^{-1}$ as $\Delta g\dop$ decreases from -0.3 eV to -0.7 eV. In  \Cref{fig7}(c), a larger $c\dopb(\infty)$ leads to a higher critical velocity.}

{In summary, the formation of SCLs at the quasi-equilibrium state has been systematically investigated under varying GB velocities and defect chemistry parameters. We conclude that
(i) higher GB velocities lead to reduced dopant segregation in the GB core, more pronounced asymmetry in the oxygen vacancy concentration profiles, increased GB potential, and an elongated SCL length in the growing grain.
(ii) the segregation energy of the dopant influences the concentration distribution pattern within the SCL and the GB potential, particularly at intermediate GB velocities. However, at extremely high GB velocities, its impact becomes negligible. (iii) the fundamental relationship between defect chemistry parameters and solute drag effects has been elucidated. When $\lvert\Delta g\V\rvert$ is smaller than 0.5 eV, the solute drag effect is negligible. However, at low $\lvert\Delta g\dop\rvert$ and dopant concentrations, minimal solute drag effects can still be observed. Thus, reducing the segregation energy of oxygen vacancy emerges as an effective strategy to minimize solute drag effects.}

\begin{figure}
    \centering
    \includegraphics[width=1.0\linewidth]{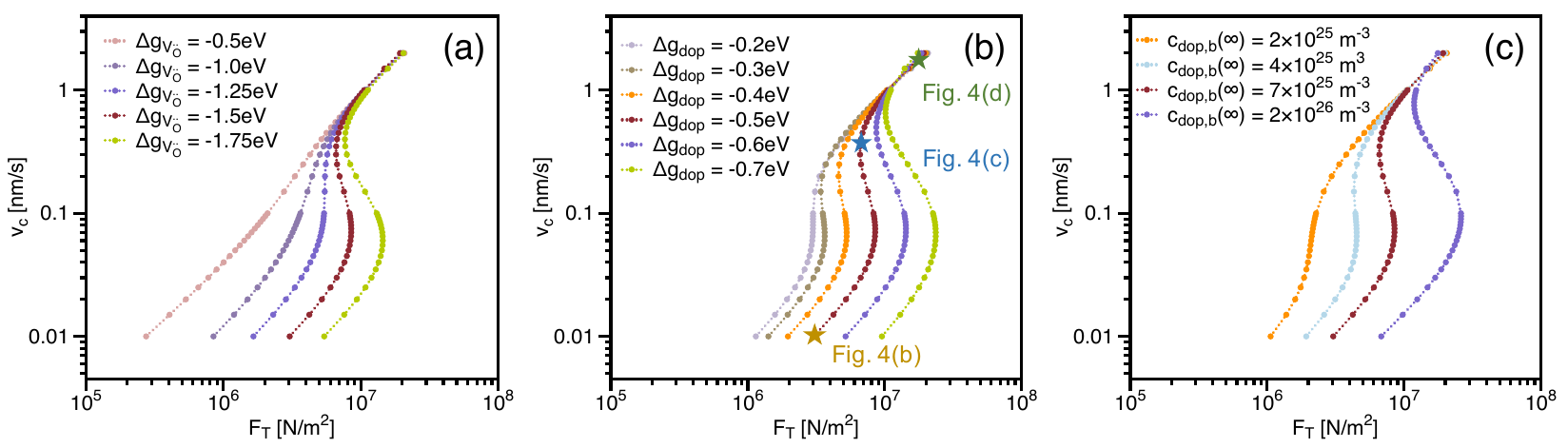}
    \caption{The total driving force on a GB with a constant velocity with {(a) $\Delta g\dop=-0.5$ eV, $c\dopb(\infty) = 7\times10^{25}$ $\text{m}^{-3} $ and different  $\Delta g\V$, (b) $\Delta g\V=-1.5$ eV, $c\dopb(\infty) = 7\times10^{25}$ $\text{m}^{-3} $  and different $\Delta g\dop$, (c) $\Delta g\V=-1.5$ eV, $\Delta g\dop=-0.5$ eV and different $c\dopb(\infty)$.} The pentagram symbols in subfigure b correspond to the different GB core velocities studied in \Cref{fig6}. The green, blue and saffron pentagram symbols indicate $v_\text{c}$=2 $\text{nm}\,\text{s}^{-1}$, 0.5 $\text{nm}\,\text{s}^{-1}$ and 0.01 $\text{nm}\,\text{s}^{-1}$, respectively. Due to the range limit of y-axis, $v_\text{c}$=0 $\text{nm}\,\text{s}^{-1}$ (\Cref{fig6} a) and 100 $\text{nm}\,\text{s}^{-1}$ (\Cref{fig6} e) are not marked. }
    \label{fig7}
\end{figure}

\subsection{Polycrystalline simulations during sintering} \label{2Dmicrostrucutre}


After systematically benchmarks and knowledge gained from quasi-static single GB cases, we investigate different grain growth patterns of materials {with only consideration of solute drag effects.}  with different defect segregation energies $\Delta g\dop$, $\Delta g\V$ and bulk defect concentrations such as $c\dopb$. 
The segregation of point defects in the core plays a significant role in influencing grain growth kinetics. To investigate the grain size distribution of STO polycrystalline material after sintering, we conduct 2D phase-field simulations. Note that three-dimensional phase-field simulations of the proposed model are achievable but require significantly more computational resources. In fact, to reveal the fundamental mechanisms, 2D simulations should be sufficient and allow for a comprehensive parameter study given the limited computational resources. 

{The defect chemistry parameters used in this section are tabulated in \Cref{quasi_Equilibrium_SCL_parameters}. Then, we describe the FEM settings for the 2D phase-field simulations as follows.  The simulation domain is 390 nm $\times$ 390 nm, with a uniform mesh spacing of $\Delta x = \Delta y = 0.39$ nm. The GB core width is set to 0.78 nm, consistent with the value used in the 1D simulations to ensure accurate reproduction of the SCLs.}
The polycrystalline microstructure is initialized using a Voronoi tessellation method, where each grain is assigned a unique grain index. The initial configuration consists of 1000 grains, and the random seed for generating the polycrystalline microstructure is set to 42. To optimize memory usage during the simulations, the grain tracker algorithm is employed {\cite{yang20193d}}.
Periodic boundary conditions are applied to all edges of the simulation domain for both the phase-field and concentration fields. For the electrostatic potential field, the bottom-left corner is grounded, while the remaining edges are set with periodic boundary conditions. {Each simulation is performed on 8 nodes (a total of 768 cores) with 3072 gigabytes of memory, requiring approximately 30 days to complete.}

{\subsubsection{Skewed grain size distribution}}
{Compared with normal grain growth pattern, the grain size distribution is skewed when the defect segregation in the GB core is considered as shown in \Cref{fig8}.} {To clearly distinguish the grain size distribution, a histogram of grain area fraction is constructed. The bin width, denoted as $\xi_\text{bin}$, is determined using the Freedman-Diaconis rule \cite{freedman1981histogram}. Subsequently, the grain size distribution is fitted as a sum of two weighted lognormal distribution functions \cite{vikrant2020electrochemicallyActa}, given by $\text{PDF}(\hat{S}) = \hat{k}_1\mathcal{N}_1(\hat{S}, \hat{\mu}_1, \hat{\sigma}_1)+\hat{k}_2\mathcal{N}_2(\hat{S}, \hat{\mu}_2, \hat{\sigma}_2)$, where $\hat{S}$ is the normalized grain area and can be expressed as  $\hat{S} = S/\langle S \rangle/\xi_\text{bin}$.  $\langle S \rangle = 152.1 $ $\text{nm}^2$ is the average area of grain at time $t=0$. $\hat{k}_1$ and $\hat{k}_2$ are the proportionality coefficients. $\hat{\mu}_1$ and $\hat{\mu}_2$ are the mean grain areas. $\hat{\sigma}_1$ and $\hat{\sigma}_2$ are the standard deviations. the lognormal distributions $\mathcal{N}_1$ and $\mathcal{N}_2$  share the general form $\mathcal{N}= \frac{1}{\hat{S}\hat\sigma\sqrt{2\pi}}\exp\left[-\frac{(\ln\hat{S} - \hat\mu)^2}{2\hat\sigma^2}\right]$.}

{The formation of {skewed grain size distribution} during the sintering process is demonstrated in \Cref{fig8}. The dopant concentration is $2\times 10^{26} $ $\text{m}^{-3}$, corresponding to a 1.1\% site fraction, with segregation energies of oxygen vacancy and acceptor dopant set to -1.5 eV and -0.5 eV, respectively.
At a sintering time of 1 h, the grain size distribution is nearly monomodal, as shown in \Cref{fig8}(a2), with only one peak observed in the fitted lognormal function. By 1.5 h, a transition from monomodal to {skewed} grain size distribution occurs, as illustrated in \Cref{fig8}(b2), where two distinct peaks emerge. {At 2 h and 3 h, the grain size distribution is skewed and do not follow log-normal distributions.}}

{The peak positions of the two grain populations are shown in \Cref{fig13}(a). For small grains, the peak position increases from 2.55 at 1.5 h to 3.07 at 3 h, while for large grains, it rises from 6.21 to 10.64 over the same period. As sintering progresses, the size disparity between small and large grains becomes increasingly significant.
In \Cref{fig13}(b), the time evolution of grain area is plotted for 10 small grains and 10 large grains. The unique grain numbers are labeled in \Cref{fig8}(d1), with black numbers indicating small grains and red numbers representing large grains. The results reveal that both small and large grains grow rapidly during the initial stage of sintering. However, as sintering progresses, the grain areas of small grains stabilize, while those of large grains continue to grow. Therefore, the growth velocity difference between two distinct grain populations leads to the formation of a {skewed grain size distribution, which do not follow the normal grain growth}.}

{Moreover, the peak positions of the two grain populations in this {skewed} grain size distribution are influenced by various defect chemistry parameters, such as the segregation energies of dopant and oxygen vacancy, as well as the dopant concentration in the bulk phase. When these defect chemistry parameters enhance the solute drag effect, smaller grains become pinned earlier and remain stable during the sintering process, leading to a more noticeable difference between the two peak positions. In the following, we examine how the selection of defect chemistry parameters affects the grain growth processes.}

\begin{figure}
    \centering
\includegraphics[width=0.9\linewidth]{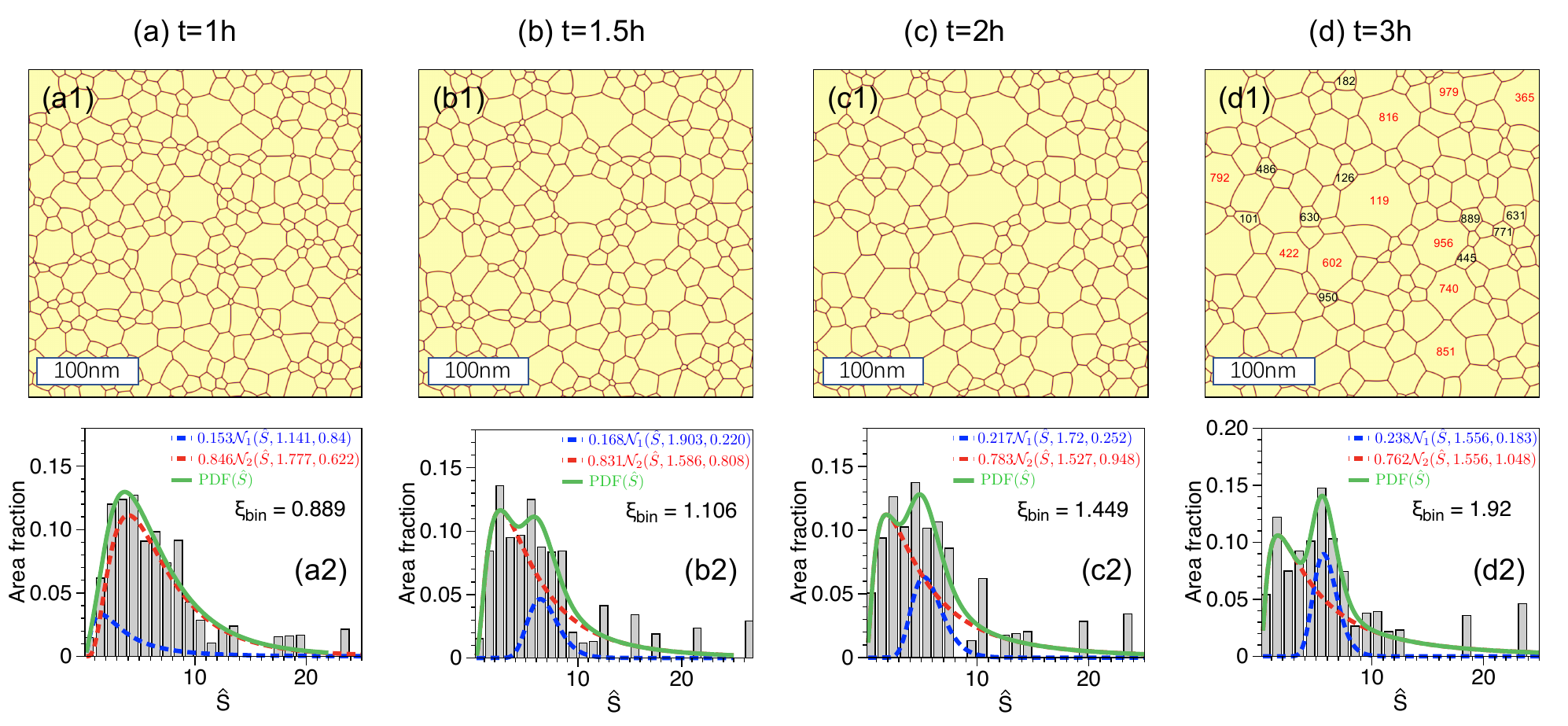}
    \caption{{The microstructure evolution during the grain growth process is presented at (a1) 1 h, (b1) 1.5 h, (c1) 2 h, and (d1) 3 h, with defect chemistry parameters set to $\Delta g\V = -1.5$ eV, $\Delta g\dop = -0.5$ eV, and $c\dopb = 2\times 10^{26} \text{m}^{-3}$. These parameters are consistent with those used in \Cref{fig9}(f).
The area fraction distribution is fitted by two lognormal distribution functions and shown in (a2)–(d2) to illustrate the {skewed} grain size distribution. The blue dashed lines represent the smaller grain population, while the red dashed lines correspond to the larger grain population. The brown solid lines indicate the combined distribution of both populations. Figure (a)  also appears in \Cref{fig9} (f) to compare with other simulations. }}
    \label{fig8}
\end{figure}

\begin{figure}
    \centering
\includegraphics[width=0.9\linewidth]{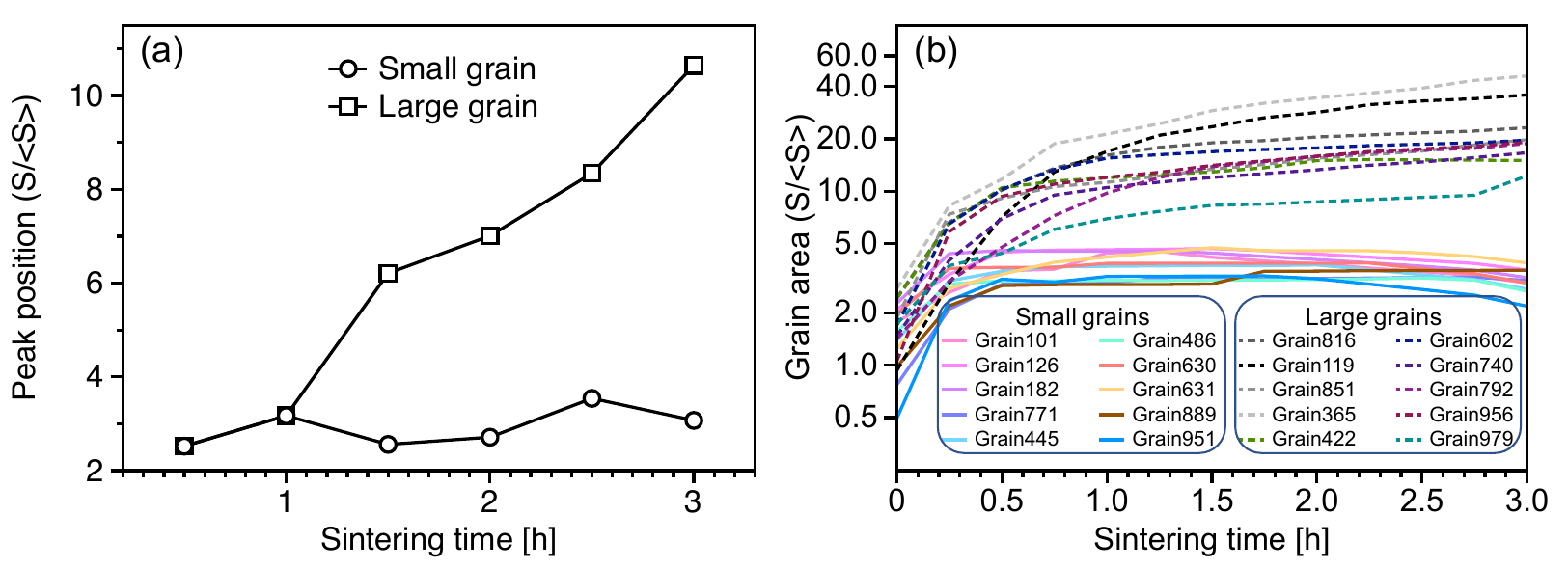}
\caption{{(a) Illustration of the grain size peak positions for two grain populations: small grains and large grains. For sintering times shorter than 1 h, a monomodal grain size distribution is observed, with only a single peak present. As sintering progresses, a {skewed} grain size distribution emerges, characterized by two distinct peaks. The difference in peak positions between small and large grains becomes increasingly pronounced with longer sintering times. (b) Time evolution of grain areas for 20 specific grains. The unique grain numbers are labeled in \Cref{fig8}. Solid lines represent small grains, while dashed lines correspond to large grains. As sintering time increases, the grain areas of large grains grow significantly, while the grain areas of small grains remain pinned.}}
    \label{fig13}
\end{figure}

\subsubsection{Influence of segregation energies and bulk defect concentration}
Subsequently, we investigate the influence of defect chemistry parameters on the microstructure evolution during sintering processes. The microstructure of grains is presented in \Cref{fig9}(a)-(g) after 1 hour of sintering at 1623 K.
In \Cref{fig9}(a), (b) and (c), we decrease the segregation energy of oxygen vacancy ($\lvert\Delta g\V \rvert$) from 1.5 eV to 0.5 eV when $\Delta g\dop$=-0.5 eV and $c\dopb = 7\times10^{25}$ $\text{m}^{-3}$. 
{Skewed} grain size distribution is clearly observed in \Cref{fig9}(a2). {The reduction of oxygen vacancy segregation energy decreases not only the segregation concentration of oxygen vacancy but also that of the dopant in the GB core. Consequently,  the solute drag effect is notably weakened. As shown in \Cref{fig9}(b2), when $\Delta g\V = -1.0 \text{eV}$, two distinct grain populations are still observed. In contrast, \Cref{fig9}(c2) shows that the grain size distribution becomes nearly monomodal when $\Delta g\V = -0.5 \text{eV}$, indicating that the solute drag effect is negligible in this case. This observation is consistent with the 1D phase-field simulation results presented in \Cref{fig7}.}
The concentration of dopant is presented in \Cref{fig9}(a3), (b3) and (c3), the formation of {solute cloud imprinted by vanished GBs} is clearly observed, especially in \Cref{fig9}(a3).
Concentration of oxygen vacancy is detailed in \Cref{fig9}(a4), (b4) and (c4). Segregation is more significant for $\Delta g\V = -1.5$ eV. Compared to the dopant, {solute cloud} of oxygen vacancies are less distinct.
In view of the higher diffusivity of oxygen vacancy, the contribution of oxygen vacancy to solute drag force is also weak. {Solute clouds} of oxygen vacancies are only observed near the junctions where grains shrink rapidly with a small radius.
The electrostatic potential distributions are depicted in \Cref{fig9}(a5), (b5) and (c5). The GB potential considerably decrease from  (a5) to  (c5). The negative potential bands can be also observed near the GB cores. As all the considered cases show, the electrostatic potentials in small shrinking grains are significantly higher than in the surrounding growing grains, leading to a stronger electrostatic contribution to the solute drag effects.  Consequently, the shrinking speed of small grains is reduced, which induces pinning effects on the surrounding growing grains. 
{Note that the heterogeneous electrostatic potential distribution arises from the velocity difference of individual GB, in accordance to the observation in the quasi-static cases in \Cref{fig6}, rather than any anisotropy which is absent in the current simulations.}

{In \Cref{fig9}(a), (d), and (e), we fix $\Delta g\V$ = -1.5 eV and $c\dopb = 7\times10^{25} $ $\text{m}^{-3}$, and examine the influence of different $\Delta g\dop$ values. A larger $\lvert \Delta g\dop \rvert$ leads to more pronounced solute drag effects. In all cases, a {skewed} grain size distribution is observed, even when $\Delta g\dop$ = -0.2 eV, as shown in \Cref{fig9}(e).
In \Cref{fig9}(d2), although the solute drag effect is more pronounced due to higher dopant segregation energy and the grain size distribution can be fitted by two lognormal distribution functions, the PDF($\hat{S}$) function still exhibits a monomodal shape. We expect that a {skewed} PDF($\hat{S}$) function that do not follow the normal grain growth will emerge if the sintering time is extended, as demonstrated in \Cref{fig8}.
Additionally, the electrostatic potential distribution remains largely unchanged, as shown in \Cref{fig9}(a5), (d5), and (e5). For the low dopant segregation energy case ($\Delta g\dop = -0.2$ eV), the solute drag effect arises primarily from the electrochemical potential gradient terms (see \Cref{drag_force1}), which is contributed by the electrostatic potential part, rather than chemical part. }

{In \Cref{fig9}(a), (f) and (g), $\Delta g\V$ and $\Delta g\dop$ are constant while $2\times10^{25} $ $ \text{m}^{-3}$, $7\times10^{25} $ $ \text{m}^{-3}$ and $2\times10^{26} $ $ \text{m}^{-3}$ are chosen for dopant concentration in bulk phase, corresponding to 0.11\%, 0.42\% and 1.1\% site fractions.  For all simulations, the solute drag effects from dopant are pronounced. The dopant clouds imprinted by vanished GBs are clearly observed behind GB cores.
In \Cref{fig9}(e2) and (g2), the {skewed} grain size distributions are presented.
However, due to the more pronounced probability of small grain, longer sintering time is required for $c\dopb(\infty)$ = $2\times10^{26} $ $ \text{m}^{-3}$ to exhibit the skewed grain size distribution that do not follow the normal grain size distribution (also see \Cref{fig8}). } 

{The influence of different defect chemistry parameters on grain growth process are elucidated in this part. First, the segregation oxygen vacancy plays a dominant role in solute drag induced {skewed grain size distribution}. Reduction of the segregation energy of oxygen vacancy is an effective way to eliminate solute drag effects and the {skewed} grain size distribution to enhance the sintering capability of oxide ceramics. In our case, when $\lvert\Delta g\V \rvert<$  0.5 eV, the {skewed} grain growth behavior can be nearly avoided. Second, the segregation energy of dopant impacts the SCL formation and GB concentration of dopant. Increase segregation energy of dopant strongly decrease the grain growth rate. However, very low $\lvert\Delta g\dop \rvert<$ still leads to {skewed grain size distribution}.  Third, dopant concentration dependent solute drag effect is also verified in the present work, even very small dopant concentration (0.11\%) can lead to the {skewed grain size distribution that do not follow log-normal distributions}.}
\begin{figure}
    \centering
\includegraphics[width=0.9\linewidth]{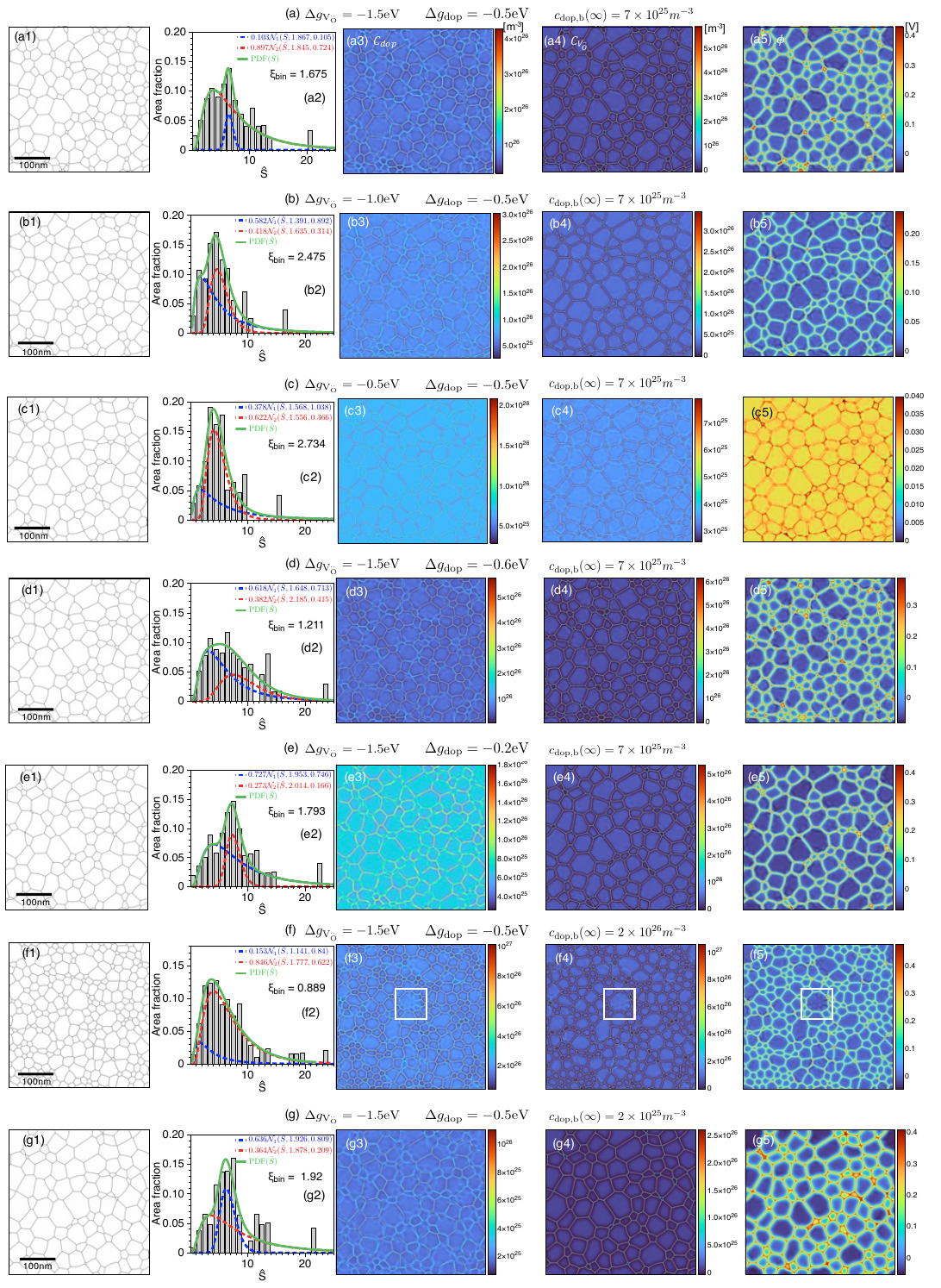}
    \caption{Phase-field simulation results of different grain growth patterns in Fe-doped STO after 1 h sintering under the influence of different defect chemistry parameters.}
    \label{fig9}
\end{figure}

\subsubsection{Grain boundary potential distribution}
{In \Cref{fig11}, we present the statistics of the average grain boundary potential in  the GB core, denoted as $\bar\phi_{\text{core}}$. The value of $\bar\phi_{\text{core}}$ is calculated as an average of the electrostatic potential along the GB core indicated by $\eta_i=0.5$ over individal GB length. \Cref{fig11}(a) illustrates the distribution of $\bar\phi_\text{core}$ for $\Delta g\V=-0.5$ eV, $\Delta g\dop=-0.5$ eV, $c\dop(\infty) = 7\times10^{25}$ $\text{m}^{-3}$ (refer to \Cref{fig9}(c) for the associated phase-field simulation result). When the solute drag effect is negligible, $\bar\phi_{\text{core}}$ is smaller than 0.05 V and approximately follows the normal distribution. In \Cref{fig11}(b), the distribution of $\bar\phi_{\text{core}}$ is presented for $\Delta g\V=-1.5$ eV, $\Delta g\dop=-0.5$ eV, $c\dop(\infty) = 7\times10^{25}$ $\text{m}^{-3}$ (refer to \Cref{fig9}(a)). Here, $\bar\phi_{\text{core}}$ significantly increases. In addition, the distribution of $\bar\phi_{\text{core}}$ changes from mono-modal to {skewed}, with one peak at 0.31 V and another at 0.35 V. In \Cref{fig11}(c), the parameters $\Delta g\V=-1.5$ eV, $\Delta g\dop=-0.5$ eV, $c\dop(\infty) = 2\times10^{26}$ $\text{m}^{-3}$ are chosen (refer to \Cref{fig9}(f)). In this case, the solute drag effect is the most pronounced in this study. Not only $\bar\phi_{\text{core}}$ increases but also the {skewed} distribution of $\bar\phi_{\text{core}}$ becomes more distinct.}
\begin{figure}
    \centering
\includegraphics[width=0.9\linewidth]{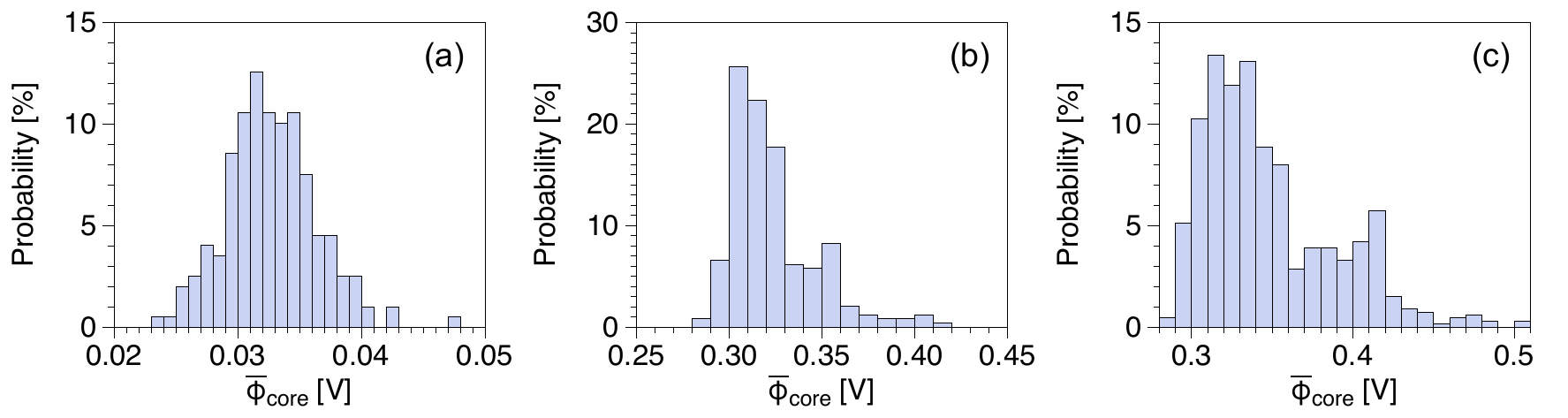}
    \caption{Probability of the average grain boundary potential in the GB cores with different defect chemistry parameters (a)$\Delta g\V=-0.5$ eV, $\Delta g\dop=-0.5$ eV, $c\dop(\infty) = 7\times10^{25}$ $ \text{m}^{-3}$ (refer to \Cref{fig9}(c)), (b)  $\Delta g\V=-1.5$ eV, $\Delta g\dop=-0.5$ eV, $c\dop(\infty) = 7\times10^{25}$ $\text{m}^{-3}$ (refer to \Cref{fig9}(a)), (c) $\Delta g\V=-1.5$ eV, $\Delta g\dop=-0.5$ eV, $c\dop(\infty) = 2\times10^{26}$ $\text{m}^{-3}$ (refer to \Cref{fig9}(f)). }
    \label{fig11}
\end{figure}

In addition, two distinct types of GBs can be observed \Cref{fig10}. The first type, characterized by a lower moving velocity, exhibits nearly symmetric concentration profiles across the GB core with a significantly higher segregation concentration of dopant. However, within a polycrystalline system, the accumulation zone of dopant at the space charge region transitions into a depletion zone.
Conversely, the second type of GB core, which moves at a higher velocity, shows lower dopant concentrations due to the break of the symmetric SCL. Similarly, the oxygen vacancy distribution in this second GB core is also asymmetric, as illustrated in \Cref{fig9}(e). \Cref{fig9}(f) presents the electrostatic potential distributions across these two types of GB cores, where the second GB core displays a larger grain boundary potential and a more pronounced electrostatic potential difference. 
{These two distinct types of GBs play different roles in the grain growth process. Fast moving GBs facilitate the formation of the large grain population during sintering, while low-moving GBs are responsible for pinning the small grain population. }

\begin{figure}
    \centering
\includegraphics[width=0.8\linewidth]{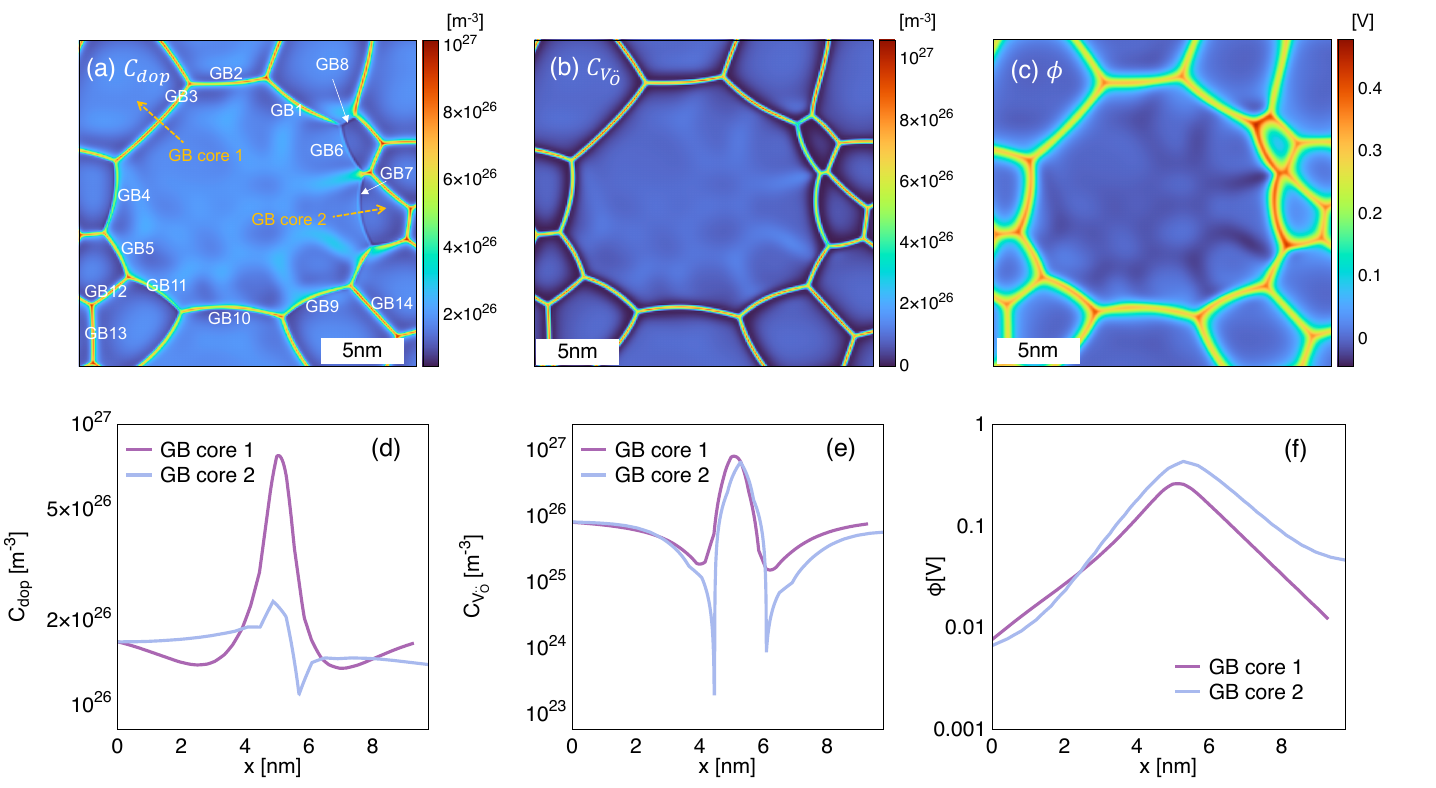}
\caption{Two types of GBs are observed in the rectangular region highlighted in \Cref{fig9}(f). The distributions of dopant concentration, oxygen vacancy concentration, and electrostatic potential in this region are shown in (a), (b), and (c), respectively. Particularly, the dopant concentration profile within the grain interior in (a) exhibits strong imprints of vanished GBs during sintering. Two types of GB cores are distinguished by dashed yellow lines in (a). The first type, characterized by a low moving velocity, has a higher segregation concentration of dopants. In contrast, the second type, with a higher core velocity, shows a lower and asymmetric dopant concentration across the GB core. Furthermore, the dopant concentration, oxygen vacancy concentration, and electrostatic potential along the two types of GB cores are plotted in (d), (e), and (f), respectively.}
    \label{fig10}
\end{figure}

{  In order to further evidence the two types of GB core, 14 different GBs are selected from the rectangular region in \Cref{fig9} (f). The relations between $v_\text{c}$ and $F_\text{T}$ as well as $\bar\phi_\text{core}$ and $F_\text{T}$ for each GB core are represented in \Cref{fig12}(b). Here, $F_\text{T}$ is the total driving force, numerically calculated via \Cref{drag_force1}. The solid lines indicate the phase-field simulation results in the bicrystal case (refer to the purple line in \Cref{fig7}(c)), while the symbols represent different GBs in the polycrystalline case.  In \Cref{fig12}(b), the relationships of $F_\text{T}$ vs. $v_\text{c}$ and  $F_\text{T}$ vs. $\bar\phi_\text{core}$ in the polycrystalline case almost follow those in the bicrystal case. There is also a velocity jump and an electrostatic potential jump when the total driving force is approximately $2\times 10^7 $ $\text{N}/\text{m}^2$. 
In the high GB velocity region, GB6, GB7 and GB8, with less dopant segregation, have much higher electrostatic potentials than the others, aligning well with the predictions in the bicrystal case. In the low GB velocity region, GB3, GB13 and GB14 also agree well with the results from the bicrystal quasi-static case. Whereas, GBs with higher velocities, such as GB2, GB4, GB11 and GB12, do not fit well into the S curve of a bicrystal case. This discrepancy can be attributed to multiple reasons. For instance, these GBs have higher GB curvatures and can interact with other GBs, which go beyond the bicystal model and can only be resolved by the grain growth simulations presented here.}

\begin{figure}
    \centering
\includegraphics[width=0.5\linewidth]{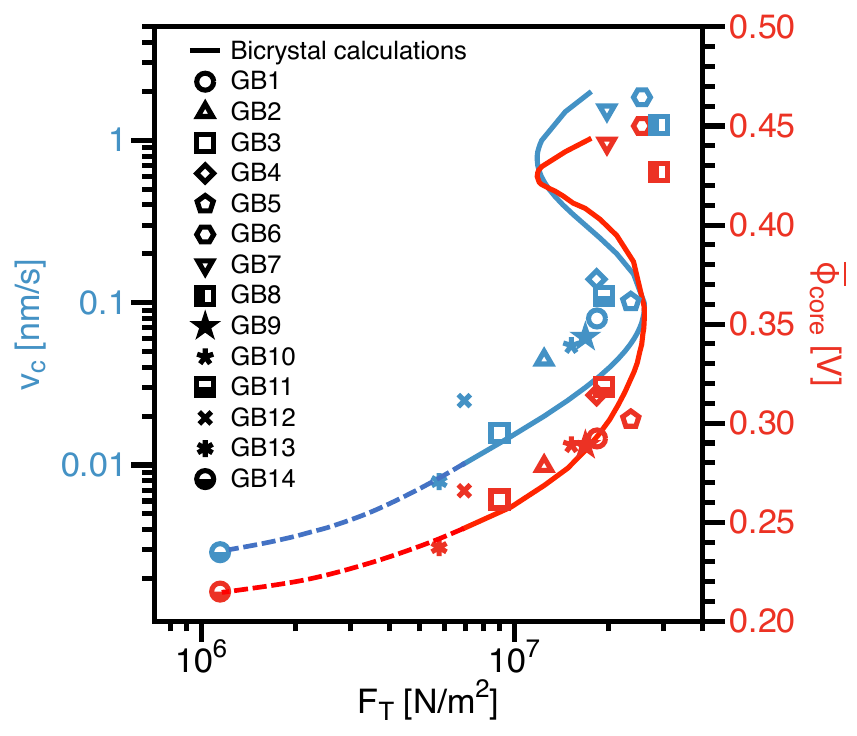}
    \caption{GB core velocity $v_\text{c}$ and average grain boundary potential $\bar\phi_\text{core}$ as functions of the total driving force $F_\text{T}$. 
    The 14 different GBs, shown in \Cref{fig10}(a), are chosen from the rectangular region in \Cref{fig9}(f) in polycrystalline simulations. The relations between $v_\text{c}$ and $F_\text{T}$ as well as $\bar\phi_\text{core}$ and $F_\text{T}$ are obtained from phase-field simulations in bicrystal and polycrystalline Fe-doped STO. Solid lines indicate the simulation results for the bicrystal case, while dashed lines are the extrapolations of the simulation results in bicystal case at very low GB velocity regime. Other symbols represent 14 different GBs chosen in the polycrystalline case.}
    \label{fig12}
\end{figure}

{In summary, we demonstrates the formation of a {skewed} grain size distribution in Fe-doped SrTiO$_3$ during the sintering process, driven solely by the solute drag effect. Two types of GBs emerge during the growth process due to defect segregation in the GB core: fast-moving GBs and slow-moving GBs. Fast-moving GBs facilitate the growth of large grains, while slow-moving GBs act to pin small grains. The coexistence of these GBs ultimately leads to {the skewed grain size distribution}.
Furthermore, variations in defect chemistry significantly influence the resulting grain size distribution. To improve the sintering process, reducing the segregation energy of oxygen vacancies proves to be an effective approach. Lowering the dopant segregation energy and bulk concentration also mitigate solute drag effects. However, even at low dopant segregation energy and bulk dopant concentration, the {skewed} grain size distribution can still occur.}

\section{Conclusion}\label{Conclusion}
In this work, we develop a phase-field model to investigate SCL formation during grain growth in oxide {electroceramics based firmly on} principles of the defect chemistry. {The main results of the paper include: }

\begin{itemize}
 \item {With the consideration of distinct segregation energies and available site densities of oxygen vacancy and acceptor dopant in bulk and GB core,} the proposed phase-field model reproduces a series of benchmark results at equilibrium of Fe-doped STO bicrystal. Simulation results, including the symmetric SCLs, grain boundary potentials, concentrations of oxygen vacancies and acceptor dopants in the GB core and the bulk phase as well as the electrostatic distributions, show good agreement with the analytical predictions in the MS and GC models.

\item Simulations are carried out comprehensively for a GB core moving with constant velocities under quasi-equilibrium state. At low core velocity, the asymmetric SCL forms. With increasing GB core velocity, the dopant segregation {diminishes}. However, the movement of oxygen vacancy within the GB core only breaks at extremely high velocity due to the high diffusivity of oxygen vacancy. In addition, the electrostatic potential in the shrinking grain becomes higher than that in the growing grain.

\item  {Skewed grain size distribution} is observed even without distinguishing any grain orientation or considering any anisotropic GB mobility in the phase-field simulations at 1623K. It indicates that the solute drag effect alone can lead to {skewed grain size distribution that do not follow the log-normal distribution in the normal grain growth}. As the formation energy differences of oxygen vacancies and acceptor dopants ($\Delta g\V$ and $\Delta g\dop$) become more negative, the grain growth velocity decreases. This trend is also observed with increasing dopant concentration ($c\dopb$). Notably, the dopant concentration profile within the grain interior distinctly reflects the significant imprint of the grain boundaries that disappeared during sintering. 

\item  Simulation results reveal two distinct types of GB cores. The first type, with nearly symmetric concentration profiles at the GB,  exhibits stronger solute drag effects, resulting in a much lower moving velocity and a tendency to remain pinned in position. The second type of GB core shows asymmetric concentration profiles and a higher moving velocity, contributing predominantly to grain coarsening. {The coexistence of these grain boundaries results in grain growth behavior that deviates from the typical normal grain growth pattern..}

\item {To improve the sintering process, reducing the segregation energy of oxygen vacancies proves to be an effective approach. Lowering the dopant segregation energy and bulk concentration also mitigate solute drag effects. However, even at low dopant segregation energy and bulk dopant concentration, the {skewed} grain size distribution can still occur.}

\item   {The skewed grain size distribution} resulting from solute drag effects leads to a large variation of grain boundary potentials. Typically, smaller grains exhibit higher grain boundary potentials at the later stages of grain growth and form more blocking GBs in term of ionic conductivity. When considering the total conductivity of an electrolyte material, the electrical current needs to find a "detour" around these blocking GBs by passing through the surrounding larger grains, with less blocking GBs. It implies that to increase the conductivity one may also optimize the microstructure so that blocking GBs can be circumvented more easily. In contrast, if blocking GBs are desired, such as in capacitors, the grain size should be kept small to prevent the development of detour paths.
\end{itemize}

The proposed phase-field model offers naturally a series of promising extensions. First, we can e.g. incorporate additional species, such as the electrons, holes and the dopants with different valance states, into the current phase-field model to quantitatively investigate the formation of the SCL. Second, GB orientation plays a crucial role in SCL formation and different grain growth patterns, and both the interface energy and the defect chemistry parameters $\Delta g\V$ and $\omega_c N\Vc$ depend on GB orientation \cite{de2019effect}.  {Third, the assumptions of dilute and non-interacting defects become invalid at high defect concentrations, where defect–defect interactions must be accounted for. Various approaches have been proposed to address this complexity. One approach involves using a regular solution model to formulate the interaction energy contribution \cite{lund2021thermodynamically, vikrant2020electrochemical, vikrant2020electrochemicallyActa}. Another approach, proposed by Mebane and De Souza (MDS) \cite{mebane2015generalised}, provides expressions for the electrochemical potentials of oxygen vacancy and acceptor dopant while incorporating defect–defect interactions. However,  MDS model does not include an explicit free energy formulation, which is essential for direct integration into phase-field models.}
Moreover, we assume in this paper only the $\text{Fe}^{3+}$ and ignore $\text{Fe}^{4+}$, which is reasonable for high temperature scenarios. But, the situation regarding solute drag effect changes at low temperature when a significant amount of $\text{Fe}^{4+}$ is present. The $\text{Fe}^{3+}$ ions in the GB core can not easily follow the movement of the GB core due to their low diffusivity. In contrast, the rapid movement of electron holes can adjust the local $\text{Fe}^{3+}$/$\text{Fe}^{4+}$ ratio and establish an equilibrium charge distribution. The solute drag effect may differ when a significant amount of $\text{Fe}^{4+}$ is considered.

\section*{Acknowledgement}
The financial support of German Science Foundation (DFG) in the framework of the Collaborative Research Centre 1548 (CRC 1548, project number 463184206) and the Project 471260201 are acknowledged. The authors K. Wang and B.-X. Xu greatly appreciate the access to the Lichtenberg II High-Performance Computer (HPC) and the technique supports from the HHLR, Technische Universit\"at Darmstadt. The computing time on the HPC is granted by the NHR4CES Resource Allocation Board under the project “special00007”.

\bibliographystyle{elsarticle-num} 
\bibliography{reference}

\end{document}